\documentclass[11pt,a4paper]{article}
\usepackage{jheppub}
\usepackage{feynmf}
\usepackage{pifont}
\usepackage{verbatim}

\newcommand{\sh}[1]{#1\hspace{-6pt}/}
\def\siml{{\ \lower-1.2pt\vbox{\hbox{\rlap{$<$}\lower6pt\vbox{\hbox{$\sim$}}}}\ }} 
\def\simg{{\ \lower-1.2pt\vbox{\hbox{\rlap{$>$}\lower6pt\vbox{\hbox{$\sim$}}}}\ }}

\title{Gauge invariant definition of the jet quenching parameter}
\author{Michael Benzke,}
\author{Nora Brambilla,}
\author{Miguel A. Escobedo,}
\author{Antonio Vairo}
\affiliation{Physik-Department, Technische Universit\"at M\"unchen, \\ James-Franck-Str. 1, 85748 Garching, Germany}
\emailAdd{michael.benzke@tum.de}
\emailAdd{nora.brambilla@ph.tum.de}
\emailAdd{miguel.escobedo@ph.tum.de}
\emailAdd{antonio.vairo@ph.tum.de}

\preprint{TUM-EFT 28/11}

\abstract{
In the framework of Soft-Collinear Effective Theory, the jet quenching parameter, $\hat{q}$, has been evaluated 
by adding the effect of Glauber gluon interactions to the propagation of a highly-energetic collinear parton
in a medium.  The result, which holds in covariant gauges, has been expressed in terms 
of the expectation value of two Wilson lines stretching along the direction of the four-momentum of the parton.
In this paper, we show how that expression can be generalized to an arbitrary gauge 
by the addition of transverse Wilson lines. The transverse Wilson lines are explicitly computed 
by resumming interactions of the parton with Glauber gluons that appear only in non-covariant gauges. 
As an application of our result, we discuss the contribution to $\hat{q}$ coming from transverse 
momenta of order $g^2T$ in a medium that is a weakly-coupled quark-gluon plasma.
}

\keywords{Jets, Heavy Ion Phenomenology}

\begin{document}

\maketitle

\section{Introduction}
There is accumulating evidence that a new state of matter, called the
quark-gluon plasma, is formed in heavy-ion collision experiments at RHIC and LHC. 
The properties of this new state can be studied by using different
observables and phenomena. One of them is the so-called jet quenching.

Jet quenching is the process by which a highly-energetic jet loses
energy while traver\-sing a medium. Jets are created in the early
stages of the collision before the plasma has been formed. 
When the parton that fragments into the jet interacts with the medium, it may lose
energy by different processes, such as gluon bremsstrahlung and pair production
\cite{Gyulassy:1993hr,Baier:1996kr,Baier:1996sk,Zakharov:1996fv,Zakharov:1997uu,Wiedemann:2000za,Arnold:2002ja,Wang:2001ifa,Gyulassy:2000er,Arnold:2008vd} (in \cite{Armesto:2011ht} there is a comparison of different formalisms and further references). 
By comparing suitable jet observables from heavy-ion collisions with observables from p-p collisions, 
it is then possible to gain information about the properties of the medium.
Jet quenching has been measured by the CMS \cite{Chatrchyan:2011sx}, 
ATLAS \cite{Aad:2010bu} and ALICE collaborations \cite{Aamodt:2010jd} at LHC, and before by the PHENIX \cite{Adcox:2001jp} 
and STAR collaborations \cite{Adler:2002xw} at RHIC. 
The measurements at RHIC have been analyzed in different theoretical models (see \cite{Bass:2008rv} for a review). 
A common feature of the models is that they need to consider the effect of the so-called transverse momentum broadening.

Transverse momentum broadening refers to a process that happens when the parton forming the jet interacts with 
constituents of the medium that have a much lower energy than the parton. 
The interaction does not change (in first approximation) the energy of the parton but it changes 
the momentum component of the parton that is transverse to the initial jet direction. 
This process is commonly characterized by the so-called jet quenching parameter, $\hat{q}$~\cite{Baier:1996sk}. 
Theoretically, $\hat{q}$ can be related in a model-independent way to   
the expectation value of two Wilson lines oriented along one 
of the light-cone directions \cite{Baier:1996kr,Zakharov:1996fv,Wiedemann:2000za,Wiedemann:2000tf,CasalderreySolana:2007zz,D'Eramo:2010ak}.
A way to derive this result is by factorizing the high-energy physics of the jet from
the low-energy physics of the medium. A convenient set up is provided by 
the Soft-Collinear Effective Theory (SCET), which is an effective field theory (EFT)
suited to describe processes involving nearly massless highly-energetic particles 
like jets~\cite{Bauer:2000ew,Bauer:2000yr,Bauer:2001ct,Bauer:2001yt,Bauer:2002nz,Beneke:2002ph}.

The expression of $\hat{q}$ in terms of Wilson lines oriented along the light-cone 
holds only in covariant gauges and is not, in general, gauge invariant.
This can best be seen by choosing the light-cone gauge $A^+(x)=0$, which 
sets all Wilson lines equal to one. In \cite{Liang:2008vz}, a similar problem was studied 
for semi-inclusive deep inelastic scattering (SIDIS); the result was extended 
by analogy to the jet quenching parameter case. The expressions obtained 
in the SIDIS case are manifestly gauge invariant, while the expression for $\hat{q}$ 
is gauge invariant only for some choice of regularization of the light-cone gauge singularity. 
In other studies, the expressions found for $\hat{q}$ are not gauge invariant;
a gauge invariant expression for $\hat{q}$ can be found in \cite{Ovanesyan:2011xy}, 
but limited to second order in the opacity expansion. 

In the last few years, some studies have addressed SCET in the light-cone gauge. 
These studies have stressed the relevance of transverse Wilson lines to restore gauge invariance 
in some observables and in the SCET Lagrangian \cite{Idilbi:2010im,GarciaEchevarria:2011md,GarciaEchevarria:2011rb}.
We will see that, as conjectured in \cite{D'Eramo:2010ak,Majumder:2012sh},  
transverse Wilson lines will also be crucial to make the expression of $\hat{q}$ gauge invariant.
Part of the difficulties in obtaining a fully gauge-invariant expression for $\hat{q}$ is related 
to the fields in the Wilson lines being path ordered but not time ordered~\cite{D'Eramo:2010ak}. 
Hence, fields at equal times located in different points of the Wilson lines are not contiguous, 
a fact that requires some care when dealing  with gauge transformations. 

The main aim of the paper is to derive a gauge invariant expression
for $\hat{q}$.  This will be achieved by explicitly calculating in
SCET and in light-cone gauge the scattering of a collinear
highly-energetic parton on a background of Glauber gluons.  Glauber
gluons are gluons that do not modify the collinear nature of the
parton, however, they can significantly change the transverse
component of its momentum.  Glauber gluons have first been included in
the SCET Lagrangian in \cite{Idilbi:2008vm}; as it will turn out, the
form of the leading-power Glauber interaction will depend on the gauge
used. Among others, Glauber gluons are responsible for the transverse
momentum broadening of the jet. Other responsibles are soft and
collinear gluons \cite{D'Eramo:2010ak}. While collinear gluons provide
an independent source of transverse momentum broadening (and energy
loss), we will argue that, at lowest order, by resumming Glauber
gluons a possible effect of soft gluons is automatically taken into
account. Another effect of soft gluons, namely the radiation of
hard-collinear gluons, will not be included in the present
analysis. More in general, we will neglect the effect of fragmentation
into (hard-)collinear partons. The fact that resumming Glauber and
soft gluons gives rise to a well-defined gauge-invariant expression
appears to justify, at least to the order at which we are working,
separating their effect from the one of (hard-)collinear gluons.

Finally, having obtained a gauge invariant expression for $\hat{q}$ will allow us to make 
use of the arguments developed in \cite{CaronHuot:2008ni,Laine:2012he} and relate, under 
some assumptions, $\hat{q}$ to thermal field averages evaluated numerically in 
lattice gauge theories. In particular, in the case of a weakly-coupled quark-gluon 
plasma of temperature $T$, we will discuss the contribution to $\hat{q}$ coming from 
transverse momenta of order $g^2T$.

The paper is organized in the following way. 
In section \ref{sec:pre}, we set the general framework of the computation.
Specifically, we will discuss the power counting of the SCET Lagrangian in different 
gauges, as well as the properties of the gauge fields at light-cone infinity. 
In section \ref{sec:jb}, we define and compute the transverse momentum broadening
of a jet and the jet quenching parameter. 
We shortly review the calculation in covariant gauges, while we detail 
the new calculation in the light-cone gauge $A^+(x)=0$.
Finally, we present an expression valid for arbitrary gauges.
As an application of our result to the case of a weakly-coupled plasma, 
in section \ref{sec:ap}, we use the gauge invariant definition of $\hat{q}$ to discuss 
the contribution from transverse momenta of order $g^2T$.
In section \ref{sec:con}, we conclude.

\section{Effective field theory, Feynman rules and power counting}
\label{sec:pre}
In this paper, we aim at obtaining a gauge invariant expression for the jet quenching parameter. 
Our work has been inspired by \cite{D'Eramo:2010ak}, 
which, in turn, follows from the study of~\cite{Idilbi:2008vm}. 
In these references, SCET was extended to include so-called Glauber gluons, which will be defined 
in the next section, and used as the framework for the calculation of jet broadening. 
This is advantageous, because in an EFT framework quantities may be calculated 
by syste\-ma\-ti\-cally expanding in the different relevant kinematical regimes already at the Lagrangian level
and with a definite power counting. Effective field theories also offer the advantage that 
some classes of potentially large perturbative contributions may be resummed 
via renormalization group equations. Here, we will make a limited use 
of the potentialities of EFTs as we will focus on few kinematical regimes 
and not solve renormalization group equations. 
Still we believe that SCET provides a transparent power 
counting and a suitable framework for further improvements.

Differently from the above references, we will assume a general gauge scenario
and explicitly work out the light-cone gauge case. This will require a different 
power counting, the introduction of new Feynman rules  and a particular care 
in treating gauge fields at infinity in one light-cone direction. 
We will discuss these issues in the rest of the section.

\subsection{Energy scales and degrees of freedom}
We are looking at highly-energetic jets propagating in a medium 
made of low-energy particles. Hence, such systems are characterized by a large 
and at least one small energy scale.
The large scale is the momentum $Q$ of the primary jet particle
(quark or gluon) in one light-cone direction (or the energy, which is
of the same order). The smaller scale is the temperature, $T$, 
or any energy scale characterizing the medium in non-thermal systems. 
This gives rise to a small dimensionless ratio $\lambda$, e.g. $\lambda=T/Q\ll 1$.

We will consider a parton moving along the light-cone direction 
$\bar n$ with initial momentum $q_0=(0,Q,0)$.\footnote{Our notation is $q=(q^+,q^-,q_\perp)$, with
$q^+=\bar n\cdot q$, $q^-= n\cdot q$, and the light-cone directions are given by 
$\bar n =\frac{1}{\sqrt{2}}(1,0,0,-1)$, $n = \frac{1}{\sqrt{2}}(1,0,0,1)$, such
that $\bar n\cdot n = 1$. The transverse momentum component is $q_\perp$.
}
Furthermore, we require the parton to travel through the medium without going
far off-shell or losing much energy through radiation; more specifically, 
the virtuality of the parton in the final state should be of the order of $Q^2 \lambda^2$ or smaller. 
Since we are looking at transverse momentum broadening through interaction with the medium, 
we further require the parton to acquire a large transverse momentum of the order of $Q\lambda$.
In short, we will consider the propagation through the medium 
of a parton whose final momentum scales like $Q(\lambda^2,1,\lambda)$; such a parton is called {\it collinear}.
The parton will eventually fragment into a jet, corresponding to a narrow cone of
particles with a large energy and a much smaller invariant mass. 

We may classify particles interacting with the parton according to the size of their momenta. 
We call {\it hard} all particles whose momentum square, i.e. virtuality, is of order $Q^2$,  
{\it soft}, particles whose momenta scale like $Q(\lambda,\lambda,\lambda)$, 
{\it Glauber}, particles whose momenta scale like $Q(\lambda^2,\lambda^2,\lambda)$ or $Q(\lambda^2,\lambda,\lambda)$ 
and {\it ultrasoft}, particles whose momenta scale like $Q(\lambda^2,\lambda^2,\lambda^2)$.\footnote{
The original definition of Glauber particles includes only particles 
with momentum scaling like $Q(\lambda^2,\lambda^2,\lambda)$. Following \cite{D'Eramo:2010ak,Ovanesyan:2011xy}, 
we extend our definition to include also particles with momentum scaling like $Q(\lambda^2,\lambda,\lambda)$ 
whose importance has been recently stressed in \cite{Qin:2012cz}.}
Therefore, we assume that particles in the medium have a typical soft momentum.

Energy loss through radiation can be induced by the emission of collinear partons.
This is the primary ingredient for the calculation of jet fragmentation.
Following analogous previous analyses, we will not consider here the effect 
due to the coupling of collinear particles, keeping however in mind that 
such an effect is not suppressed by any power counting.
A first study  including collinear radiation effects on the propagation of 
an energetic parton in a medium can be found in \cite{D'Eramo:2011zz}.

If coupled to a collinear particle, soft modes give rise to a so-called {\it hard-collinear} 
particle with momentum $Q(\lambda,1,\lambda)$, which is off shell by a momentum square of order $Q^2\lambda$. 
It has been argued in \cite{D'Eramo:2010ak} that because the virtuality of this particle 
is larger than the one of a collinear particle, which is $Q^2 \lambda^2$,
the effect due to the interaction with soft modes is suppressed in the strong coupling constant.
In this work, we will, however, explicitly include soft and hard-collinear modes 
and show that their possible effect can be cast in the same gauge-invariant expression 
that encompasses the effect of Glauber gluons.

The most relevant contribution to the transverse momentum broadening 
of a single collinear parton originates from the interaction with Glauber gluons. 
In the following, we will consider their contribution together with the one of soft gluons.

\subsection{SCET}
We restrict our analysis to the case of a highly-energetic primary parton that is a light quark. 
The SCET Lagrangian that describes the propagation of a massless quark in the light-cone 
direction $\bar{n}$ is 
\begin{equation}
\mathcal{L}_{\bar{n}}=\bar{\xi}_{\bar{n}}\,i\sh{n}\,\bar{n}\cdot D\,\xi_{\bar{n}}
+\bar{\xi}_{\bar{n}}\,i\sh{D}_\perp\,\frac{1}{2in\cdot D}\,i\sh{D}_\perp\,\sh{n}\,\xi_{\bar{n}}\,,
\label{eq:QCD}
\end{equation}
where $\xi_{\bar{n}}$ denotes the quark field and $iD_\mu=i\partial_\mu+gA_\mu$.
The Lagrangian \eqref{eq:QCD} is suited to describe the propagation of
a collinear quark interacting with soft, collinear, Glauber or
ultrasoft gluons. Hence, the quark field $\xi_{\bar{n}}$ is collinear
when acting on the initial and final states, but may describe both
hard-collinear and collinear virtual quarks.  A theory describing
hard-collinear and collinear quarks interacting with soft gluons has
been called SCET(hc,c,s) in the literature \cite{Beneke:2003pa}. If
the physical process of interest only contains collinear external
particles, it is possible to integrate out the hard-collinear modes to
obtain SCET(c,s), also known as SCET$_\mathrm{II}$
\cite{Bauer:2002aj}. In this work, however, we will not perform this
second step but instead consider the hard-collinear modes
explicitly. Of course, the results are the same in both approaches, if
hard-collinear quarks appear only as internal lines.

We now proceed through the following two steps.
First, we rescale the collinear field  $\xi_{\bar{n}}$ by the large momentum component 
$Q$, according to $\displaystyle \xi_{\bar{n}} \to e^{-iQx^+}\xi_{\bar{n}}$. 
This implies that the residual momentum of the quark along the light-cone 
direction $\bar{n}$, which is due to the interaction with the medium, is now of order $Q\lambda$ or smaller.
Second, we exclude from the Lagrangian collinear gluons.
As we argued in the previous section, this last requirement does not rely on the power 
counting, but on the convenience to split the calculation in a part 
that deals with one collinear particle in the final state and a part  
that deals with more than one collinear particle in the final state, 
while postponing the latter for future considerations.
Moreover we exclude possible hard-collinear radiation of gluons scaling 
like $Q(\lambda,1,\sqrt{\lambda})$ \cite{Idilbi:2008vm}, which appears also 
as a leading-order effect, when considering the interaction with soft gluons.
Under the above conditions, we have $2in\cdot D = 2Q + \hbox{residual momenta of order\;} Q\lambda$
or smaller. This implies that the last operator in \eqref{eq:QCD} may be expanded 
into local operators, which leads to\footnote{We have used
$$
\sh{D}_\perp\sh{D}_\perp = -D_\perp\cdot D_\perp - \frac{i}{2}gF^{\mu\nu}_\perp\gamma_\mu\gamma_\nu\,,
$$
where $F^{\mu\nu}_\perp = i[D^{\mu}_\perp,D^{\nu}_\perp]/g$ is the  gluon field strength.
Here and in the following, $v_\perp$ denotes a vector such that  
$v_\perp^\mu = (0,v^1,v^2,0)$ and 
$v_\perp^2 = \displaystyle (v^1)^2 + (v^2)^2 = -v_\perp^\mu v_{\perp\,\mu}$.
}
\begin{equation}
\mathcal{L}_{\bar{n}}=\bar{\xi}_{\bar{n}}\,i\sh{n}\,\bar{n}\cdot D\,\xi_{\bar{n}}
+\bar{\xi}_{\bar{n}}\,\frac{D_\perp^2}{2Q}\,\sh{n}\,\xi_{\bar{n}}
+\bar{\xi}_{\bar{n}}\,i\frac{gF^{\mu\nu}_\perp}{4Q}\,\gamma_\mu\gamma_\nu\,\sh{n}\,\xi_{\bar{n}} + \dots \;,
\label{eq:SCET}
\end{equation}
where the dots stand for higher-order terms in the $\lambda$ expansion.
The SCET Lagrangian \eqref{eq:SCET} amounts to having integrated out from QCD 
hard modes at leading order in the strong coupling constant, as well as 
the small components of the collinear fields.

The SCET Lagrangian that describes the propagation of a collinear 
particle and its interaction with Glauber gluons has been introduced in the context of jet scattering
on cold nuclear matter in \cite{Idilbi:2008vm}; Glauber gluons have also 
been found necessary for the consistency of factorization proofs in certain
processes like Drell--Yan \cite{Bauer:2010cc}.

\subsection{Power counting and leading-order Lagrangian in covariant gauge}
\label{sec:pccov}
In the SCET Lagrangian, collinear and hard-collinear quark fields,
$\xi_{\bar{n}}(x)$, scale in the same way.
This follows from the fact that both the inverse propagator in momentum space 
of a hard-collinear quark and the typical four-momentum region occupied 
by a hard-collinear quark are enhanced by the same factor, $\lambda$, 
with respect to the collinear quark case; hence, hard-collinear and collinear 
quark propagators scale in the same manner in position space and so do 
hard-collinear and collinear quark fields.
The operators $\bar{n}\cdot \partial$ and $\nabla_\perp^i=\partial_{\perp\,i}$ 
scale like $Q\lambda^2$ and $Q\lambda$ respectively when acting on a collinear field $\xi_{\bar{n}}(x)$, 
and both scale like $Q\lambda$ when acting on a hard-collinear field $\xi_{\bar{n}}(x)$. 
Soft-gluon fields scale like $Q\lambda$ and ultrasoft-gluon fields scale like $Q\lambda^2$,  
for they are homogeneous in the soft and ultrasoft scale respectively. 
In contrast, the power counting of Glauber gluons depends on the gauge.
The equations of motion require $A^+(x)$ to scale like $\bar{n}\cdot\partial$ when acting on a 
collinear quark, or smaller. We assume $A^+(x) \sim Q\lambda^2$.
In a covariant gauge, if the gluon field is coupled to a homogeneous soft source, 
this also implies $A_\perp(x) \sim Q\lambda^2$. 
In \cite{Idilbi:2008vm,Ovanesyan:2011xy} other sources have been considered that lead to 
different power countings.\footnote{
Soft sources have been considered both in \cite{D'Eramo:2010ak} and \cite{Ovanesyan:2011xy}.
In \cite{D'Eramo:2010ak}, for Glauber gluon momenta scaling like $Q(\lambda^2,\lambda^2,\lambda)$
they find that all components of the Glauber field scale like $Q\lambda^2$, 
while in  \cite{Ovanesyan:2011xy} for Glauber gluon momenta scaling like $Q(\lambda^2,\lambda,\lambda)$
they find that all components of the Glauber field scale like $Q\lambda$. 
Both results, however, seem to follow from an incorrect handling 
of the phase-space integral $\displaystyle \int d^4p \, \delta(p^0-E)$ where $E$ generally scales like 
$Q\lambda$, which is the typical energy of the soft source. The integral, correctly e\-va\-lu\-ated, vanishes over the 
momentum region $p\sim Q(\lambda^2,\lambda^2,\lambda)$, for $p^0 \sim Q\lambda^2 \ll E \sim Q\lambda$, 
while it is $\displaystyle \int d^4p \, \delta(p^0-E) = \int dp^+\,dp^-\,d^2p_\perp \, \delta(p^+/\sqrt{2}+p^-/\sqrt{2}-E)
= \sqrt{2}\int dp^+\,d^2p_\perp \sim Q^3\lambda^4$ over the momentum region $p\sim Q(\lambda^2,\lambda,\lambda)$. 
Hence it is only Glauber gluons whose momenta scale like $Q(\lambda^2,\lambda,\lambda)$
that couple at leading power to a soft source. Using the above integral, it follows that all components 
of Glauber fields of this kind scale like $Q\lambda^2$, which is the scaling adopted in this paper.
The identification of  $\displaystyle \int d^4p \, \delta(p^0-E)$ with $\displaystyle \int d^3p$ 
is correct only for Glauber gluons whose momentum scales like 
$Q(\lambda^2,\lambda,\lambda)$ and after replacing $p^3$ by the small momentum 
$\sqrt{2}\,p^+$. A similar observation can be found in~\cite{Idilbi:2008vm}.
}

According to the power counting, the leading-order SCET Lagrangian in a covariant gauge is 
\begin{equation}
\mathcal{L}_{\bar{n}}^{\rm LO,\, cov}=\bar{\xi}_{\bar{n}}\,i\sh{n}\,\bar{n}\cdot D\,\xi_{\bar{n}}
+\bar{\xi}_{\bar{n}}\,\frac{\nabla_\perp^2}{2Q}\,\sh{n}\,\xi_{\bar{n}}
\,.
\label{eq:SCET:cov}
\end{equation}
The first term on the right-hand side may either involve the interaction of two collinear quarks 
and a Glauber gluon or the interaction of a collinear quark with a soft gluon 
and a hard-collinear quark. Both interactions are of order one:
$\displaystyle \int d^4x\,\bar{\xi}_{\bar{n}}\,A^+(x)\,\xi_{\bar{n}} \sim 1$. This implies 
that a collinear quark may exchange in the medium an arbitrary number of Glauber or 
soft gluons, all contributing to the same order in $\lambda$ to the scattering amplitude.
The second term on the right-hand side of eq. \eqref{eq:SCET:cov}, instead, involves only 
collinear quarks, the hard-collinear terms being suppressed by one power of $\lambda$. 
From the leading-order Lagrangian, it follows that only the propagator of a collinear quark, 
of a hard-collinear quark  and the interaction vertex with the $A^+$ component of a gluon matter. 
The propagator of a collinear quark carrying momentum $q$ is
\begin{equation}
\frac{iQ}{2Qq^+-q_\perp^2+i\epsilon}\sh{\bar{n}}\,,
\label{collprop}
\end{equation}
the propagator of a hard-collinear quark carrying momentum $q$ is
\begin{equation}
\frac{i}{2q^++i\epsilon}\sh{\bar{n}}\,,
\label{hcollprop}
\end{equation}
whereas the vertex of the $n_\mu A^{+\,a}$ component of a gluon with a collinear or hard-collinear quark is
\begin{equation}
igT^a\bar{n}^\mu\sh{n}\,.
\label{vertexcov}
\end{equation}

The Lagrangian \eqref{eq:SCET:cov} contains, in principle, also ultrasoft gluons.
It has been shown, however, that ultrasoft gluons decouple at lowest order from collinear quarks trough the field redefinition
$\displaystyle \xi_{\bar{n}}(x) \to {\rm P}\,\exp\,\left[ig \int_{-\infty}^{x^-}dy\, \bar{n}\cdot A_{\rm us}(x^+,y,x_\perp)\right]\xi_{\bar{n}}(x)$, 
where ${\rm P}$ stands for the path ordering operator and $A_{\rm us}$ for an ultrasoft gluon field \cite{Bauer:2001yt}.
The field redefinition works for ultrasoft gluons because their transverse momentum 
is suppressed with respect to the transverse momentum of collinear quarks, 
so that, at lowest order, the kinetic energy operator, $\nabla_\perp^2/(2Q)$, commutes with ultrasoft gluons. 
Note that, for the opposite reason (the transverse momentum of Glauber gluons is of the same order 
as the transverse momentum of collinear quarks), the field redefinition would not decouple, 
even at lowest order, Glauber gluons from collinear quarks.

We conclude this section by observing that in a covariant gauge the free gluon propagator 
of a Glauber gluon may be approximated by 
\begin{equation}
\int d^4x\, e^{ik\cdot x}\langle 0|\, {T}\left(A^\mu(x)A^\nu(0)\right)|0\rangle \approx 
-\frac{i}{k_\perp^2}\left[ g^{\mu\nu} -\alpha\frac{(k_\perp^\mu+\bar{n}^\mu k^-)(k_\perp^\nu+\bar{n}^\nu k^-)}{k_\perp^2} \right] \,,
\label{eqn:Gprop}
\end{equation}
where $\alpha$ is a gauge parameter.
Equation \eqref{eqn:Gprop} follows from $k^2 \approx k_\perp^2 \sim Q^2\lambda^2$ 
and $k^{+} \sim Q\lambda^2 \ll k_\perp \sim Q\lambda$.
Hence, in the case of Glauber gluons and at lowest order in $\lambda$, 
the gluon propagator does not depend on the light-cone momentum component $k^+$.
It still can transfer such a momentum to a source coupled to the propagator. 
However, if the source is soft, i.e. it supports only soft momenta, the change to the light-cone $n$-component 
of the source momenta due to the Glauber interaction can be neglected. 
This can also be understood as follows. When a Glauber field connects to a soft source one can write
\begin{equation}
A^\mu(x) = \int d^4y \int\frac{d^4k}{(2\pi)^4}\, D^{\mu\nu}(k)\, e^{-ik\cdot (x-y)}\, J_\nu(y)\,,
\label{eqn:sSource}
\end{equation}
where $D^{\mu\nu}(k)$ is given by eq.~(\ref{eqn:Gprop}) and $J_\nu(y)$ is the source. 
Because the source is soft, it supports $y^-$ only in the region $y^- \sim 1/\lambda$.
Since $k^+$ is a Glauber momentum, the exponent $k^+y^-$ is therefore suppressed, 
which implies that the Fourier transform of the source is insensitive to the value of $k^+$ at leading order.
Note that the argument is general and holds for any soft source $J_\nu(y)$. 
For instance, in non-abelian theories it may contain
derivatives with respect to $y$, e.g. from the triple-gluon vertex. 
Because the $n$-component of such a derivative yielding a 
$k^+$-dependent term in the integral of eq. \eqref{eqn:sSource}  
contributes to the gauge field at order $Q\lambda^3$, 
it holds also in this case that the leading-order contribution 
to the gauge field comes from terms that are independent of $k^+$.

\subsection{Light-cone gauge}
The Lagrangian \eqref{eq:SCET:cov} cannot be the lowest-order Lagrangian in the light-cone 
gauge, $A^+(x) = 0$, because, otherwise, the collinear quarks would decouple from the gluons and propagate 
freely. It turns out, indeed, that in light-cone gauge, Glauber gluons scale differently than 
in covariant gauges and, because of this, new interaction vertices 
between collinear quarks and gluons must be considered at lowest order.

The different scaling of the gluon fields in the light-cone gauge $A^+(x)=0$ 
is due to the fact that, in that gauge, the gluon propagator is singular for $k^+\rightarrow 0$. 
As a consequence, $A_\perp(x)$ does not vanish for $x^-=\pm \infty$, like in covariant gauges, 
and may be conveniently decomposed into \cite{Idilbi:2010im,GarciaEchevarria:2011md} 
\begin{eqnarray}
&& A_\perp^i(x) = A^{\mathrm{cov},i}_\perp(x) + A^{\mathrm{sin},i}_\perp(x)\,,
\label{eqn:gluonDecomp0}
\end{eqnarray}
with
\begin{eqnarray}
&& A^{\mathrm{sin},i}_\perp(x) = \theta(x^-)A_\perp^i(x^+,\infty,x_\perp)+\theta(-x^-)A_\perp^i(x^+,-\infty,x_\perp)\,.
\label{eqn:gluonDecomp}
\end{eqnarray}
The field $A^{\mathrm{cov},i}_\perp(x)$ contributes to the non-singular part of the propagator and 
vanishes at $x^-=\pm\infty$, while the field $A^{\mathrm{sin},i}_\perp(x)$ contributes to the singular part of the propagator.
Note that the theta functions, $\theta(\pm x^-)$, are singular in momentum space for $k^+\rightarrow 0$.
They multiply (non-vanishing) fields evaluated at $x^-=\pm\infty$. 

Before turning to the SCET Lagrangian, it is useful to consider some 
general properties of gauge fields at $x^-=\pm\infty$ \cite{Belitsky:2002sm}. 
The energy of the gauge field is proportional to the integral over space of ${\bf E}^2+{\bf B}^2$, 
where ${\bf E}$ and ${\bf B}$ are the chromoelectric and chromomagnetic fields respectively. 
The energy is finite if both fields vanish at infinity. 
This, in turn, requires the fields $A_\mu$ to be pure gauge at infinity; 
so that at infinity we may write
\begin{equation}
A_\mu(x)=\frac{i}{g} \,\partial_\mu \Omega(x)\, \Omega^\dagger(x)\,,
\end{equation}
where $\Omega(x)$ is an SU(3) gauge transformation.  For an infinitesimal gauge 
transformation $\Omega(x)\approx 1-ig\phi(x)$, this implies  
\begin{equation}
A_\mu(x)=\partial_\mu\phi(x)\,.
\label{gaugeinfinity}
\end{equation}

In the light-cone gauge $A^+(x)=0$, the condition at infinity \eqref{gaugeinfinity} translates 
into the following condition for the transverse components of the gauge fields at $x^-=\pm\infty$:
\begin{equation}
A_\perp^i(x^+,\pm\infty,x_\perp) = - \nabla_\perp^i\phi^\pm(x^+,x_\perp)\,.
\label{eqn:puregauge}
\end{equation}
The fields $\phi^\pm$, solution of eq. \eqref{eqn:puregauge}, are
\begin{equation}
\phi^\pm(x^+,x_\perp) = - \int_{-\infty}^0 ds\, l_\perp \cdot A_\perp(x^+,\pm\infty,x_\perp+l_\perp s)\,,
\label{eqn:phi}
\end{equation}
where, for convenience, we have chosen to integrate over a straight line going from 
$x_\perp-\infty l_\perp$ to $x_\perp$,  $l_\perp$  being an arbitrary 
vector in the transverse plane. The solution does not depend on 
the chosen integration path: for a path variation can be written as  
a surface integral over a field strength tensor \cite{Migdal:1984gj}, 
which vanishes at $x^-=\pm\infty$. 
The solution does depend instead on the starting point of the integration path, 
which, in our case, has been chosen to be $x_\perp-\infty l_\perp$.
In a covariant gauge, because there is no preferred direction in space, 
this cannot happen, which is another 
way to see why the field has to vanish at $x^-=\pm\infty$. 
In light-cone gauge, different starting points lead to different fields $\phi$
that are related by a gauge transformation.  
Gauge invariance then requires physical observables, like the jet quenching 
parameter, to be independent of the choice of the starting point.
We will explicitly show that this is indeed the case.

\subsection{Power counting and leading-order Lagrangian in light-cone gauge}
\label{sec:pclc}
In the light-cone gauge $A^+(x)=0$ and for soft sources, 
the Glauber fields $A^-(x)$ and $A_\perp^{\mathrm{cov}}(x)$ scale 
as in the covariant gauge case, i.e. like $Q\lambda^2$. 
However, the singular part of the Glauber field $A_\perp(x)$ is enhanced by one power of $1/\lambda$,
so that we have  $A^{\mathrm{sin},i}_\perp(x) \sim Q\lambda$.\footnote{
This may be traced back to the factor $k^i_\perp/k^+$ appearing in the Fourier transform of the gluon 
propagator $\langle 0| \, {T}\left(A^{\mathrm{sin}}_\perp(x)A^{\mathrm{sin}\,-}(0)\right)|0\rangle$;  
$k^i_\perp/k^+$ is of order $1/\lambda$ for Glauber gluons.}
Soft and ultrasoft gluons scale like in the covariant gauge case and do not contribute at lowest-order 
to the Lagrangian $\mathcal{L}_{\bar{n}}$.
From eq.~\eqref{eq:SCET} and according to the above power counting, the leading-order SCET Lagrangian 
in light-cone gauge is then 
\begin{equation}
\mathcal{L}_{\bar{n}}^{\rm LO,\, lc}=\bar{\xi}_{\bar{n}}\,i\sh{n}\,\bar{n}\cdot\partial\,\xi_{\bar{n}}
+\bar{\xi}_{\bar{n}}\,\frac{(\nabla_\perp + ig A^{\mathrm{sin}}_\perp)^2}{2Q}\,\sh{n}\,\xi_{\bar{n}}\,.
\label{eq:SCET:lc}
\end{equation}
The gluons in the Lagrangian are Glauber gluons and the quark fields are collinear quark fields.
Note that the operator proportional to $F^{\mu\nu}_\perp$ does not 
appear at leading order because the fields $\phi^\pm$ do not contribute to the 
field strength tensor, which, in turn, reflects the fact that the field strength tensor 
vanishes at infinity.

According to the leading-order Lagrangian \eqref{eq:SCET:lc},
in light-cone gauge, collinear quarks couple to (transverse) gluons in two possible ways: 
via a 1-gluon and a 2-gluon vertex. 
Keeping Glauber gluons as background fields, the Feynman rule for the 1-gluon vertex reads
\begin{equation}
\raisebox{-30pt}{\includegraphics[width=35mm]{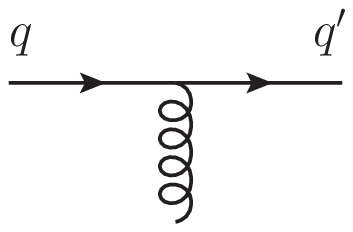}}=
-i\,\frac{q'_\perp\cdot gA_\perp^{\mathrm{sin}}(q'-q)+gA_\perp^{\mathrm{sin}}(q'-q)\cdot q_\perp}{2Q}\,\sh{n}\,,
\label{eq:feyn1}
\end{equation}
whereas the Feynman rule for the 2-gluon vertex is 
\begin{equation}
\raisebox{-30pt}{\includegraphics[width=35mm]{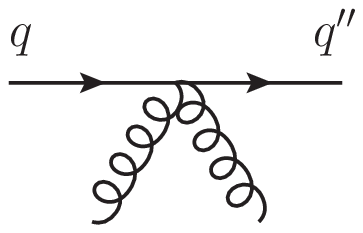}}
=-\frac{i}{2Q}\int\frac{d^4q'}{(2\pi)^4}\,gA_\perp^{\mathrm{sin}\,i}(q''-q')\,gA_\perp^{\mathrm{sin}\,i}(q'-q)\,\sh{n}\,.
\label{eq:feyn2}
\end{equation}

\section{Jet broadening}
\label{sec:jb}
The jet quenching parameter, $\hat{q}$, can be defined as the average square
transverse momentum with respect to the original direction of motion that 
a highly-energetic parton picks up while travelling a large distance $L$ through the medium. 
Following the notation and the derivation of \cite{D'Eramo:2010ak}, $\hat{q}$ can be written as
\begin{equation}
\hat{q} = \frac{1}{L}\langle k_\perp^2\rangle = \frac{1}{L}\int\frac{d^2k_\perp}{(2\pi)^2}\,k_\perp^2 P(k_\perp)\,,
\label{eq:qhat}
\end{equation}
where $P(k_\perp)$ is the probability for the hard parton to pick up a certain transverse momentum $k_\perp$. 
The calculation of $P(k_\perp)$ and $\hat{q}$ proceeds then through the following steps.

\medskip

\begin{figure}[ht]
\center
\includegraphics[width=100mm]{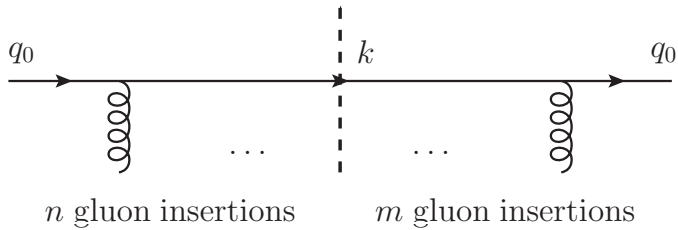}
\caption{Cutting diagram contributing to ${\cal A}_{mn}$. The dashed line is the cut.
The continuous line across the cut in the initial state and in the final state represents a collinear parton. 
}
\label{fig:Amn}
\end{figure}

(1) First, we consider a highly-energetic parton of momentum $q_0=(0,Q,0)$ in an initial state, $|\mathrm{in}\rangle$,
propagating in a box of length $L$ (volume $L^3$) and interacting with an arbitrary number of background gluons 
from the medium leading to a final state, $|k,\sigma\rangle$, made of a collinear parton of momentum $k$ and 
polarization $\sigma$. Squaring the amplitude and integrating/summing over the final state momenta/polarizations and colors, 
we obtain 
\begin{eqnarray}
{\cal A} = \frac{1}{\sqrt{2}QL^3}\,\sum_{\sigma}\,\int\frac{d^3k}{(2\pi)^3\,2|{\bf k}|}\; 
\left| \langle k,\sigma|\,{\rm T}\,|\mathrm{in}\rangle \right|^2\,,
\end{eqnarray}
where the initial state has been normalized relativistically 
(the initial state energy is $Q/\sqrt{2}$) in a box of size $L$.
T is the interaction part of the S-matrix. It is useful to write  
${\rm T} = \sum_n {\rm T}_n$, $n$ being the number of gluon lines 
attached to the parton, and define 
\begin{eqnarray}
\frac{d^2{\cal A}_{mn}}{dk_\perp^2}
=  \frac{1}{\sqrt{2}QL^3}\,\sum_{\sigma}\,\int\frac{dk^3}{(2\pi)\,2|{\bf k}|}\; 
\langle\mathrm{in}|\,{\rm T}^\dagger_m\,|k,\sigma\rangle \langle k,\sigma|\,{\rm T}_n\,|\mathrm{in}\rangle\,,
\label{eqn:amp}
\end{eqnarray}
so that 
\begin{equation}
{\cal A} = \sum_ {mn} \int \frac{d^2k_\perp}{(2\pi)^2}\; \frac{d^2{\cal A}_{mn}}{dk_\perp^2}\,.
\end{equation}
A convenient way to calculate ${\cal A}$ is via the optical theorem.
This amounts to computing twice the imaginary part of all diagrams made 
of an incoming and outcoming parton of momentum $q_0$ interacting 
with an arbitrary number of background gluons. The quantity ${\cal A}_{mn}$ is then 
the cutting diagram made of $n$ gluons before the cut and $m$ gluons after the cut, see fig.~\ref{fig:Amn}.
In the quark case, cutting the collinear quark propagator  
amounts to picking up the discontinuity of eq.~\eqref{collprop}, 
while hard-collinear propagators do not contribute to the cut 
because they do not correspond to on-shell modes.
Vertices and propagators on the left-hand side of the cut are those 
defined in sections \ref{sec:pccov} or \ref{sec:pclc}; vertices and 
propagators on the right-hand side of the cut are the complex conjugated 
of those on the left-hand side. Specifically the $i\epsilon$ prescription 
of the propagators on the right-hand side comes with an opposite sign.

\medskip

(2) The probability $P(k_\perp)$ is related to $d^2{\cal A}_{mn}/dk_\perp^2$ by the equations:
\begin{eqnarray}
P(k_\perp) = \left\{
\begin{array}{ll}
\displaystyle
\frac{L}{\Delta t} \sum_{mn}\left\langle \frac{d^2\bar{\cal A}_{mn}}{dk_\perp^2}\right\rangle 
&\hspace{4mm} \hbox{for}\; k_\perp\neq 0 \\
\displaystyle
\frac{L}{\Delta t} \sum_{mn}\left\langle \frac{d^2\bar{\cal A}_{mn}}{dk_\perp^2}\right\rangle 
- \frac{L}{\Delta t} \sum_{mn} L^2 \int \frac{d^2k^\prime_\perp}{(2\pi)^2}\;
\left\langle \frac{d^2\bar{\cal A}_{mn}}{dk^{\prime\,2}_\perp}\right\rangle + L^2
&\hspace{4mm} \hbox{for}\; k_\perp = 0 
\end{array}
\right.
\label{Pkperp}
\end{eqnarray}
where $\bar{\cal A}_{mn}$ stands for ${\cal A}_{mn}$ averaged over the initial-state 
polarizations and colors, the brackets, $\langle \cdots \rangle$,  denote a field average (e.g. a 
thermal field average) and $\Delta t$ is the emission time.\footnote{
As $L$ provides the length traveled by the parton in the medium, $\Delta t$ provides the time.
In particular, we have 
$$
\int_{-\frac{\Delta t+L}{2\sqrt{2}}}^{\frac{\Delta t+L}{2\sqrt{2}}}dx^+ =  \frac{\Delta t + L}{\sqrt{2}}\,,
$$
which is about $\Delta t/\sqrt{2}$ for $\Delta t \gg L$. The probability $P(k_\perp)$ is then 
related to an amplitude square normalized with respect to the number of particles 
propagating through the medium, i.e. $\Delta t/L$. 
} Eventually we take the limit $L\to\infty$ that corresponds to an infinite medium.

\medskip

(3) Finally, the jet quenching parameter is obtained by integrating over the probability  $P(k_\perp)$ as in 
eq. \eqref{eq:qhat}. Since in our computation we are considering contributions 
relevant for a transverse momentum broadening of the parton of order $Q\lambda$ or softer, 
the integral in the transverse momentum cannot exceed a cut-off $q^{\rm max}$, which is 
$Q\lambda \siml q^{\rm max} \ll Q$.

\subsection{Jet broadening in covariant gauge}
\label{sec:Pcg}
The calculation in SCET and in a covariant gauge of $P(k_\perp)$ has been done in \cite{D'Eramo:2010ak}.
Here we sketch a partially different derivation.

In a covariant gauge, the diagram of fig.~\ref{fig:Amn} involves only vertices of the type \eqref{vertexcov} 
that couple collinear quarks with $A^+$ fields of Glauber or soft gluons. 
Let us first consider the case of Glauber gluons only; the diagram is proportional to 
\begin{equation}
\int \prod_i \frac{d^4q_i}{(2\pi)^4} \,\cdots\, 
\frac{iQ}{2Qq_2^+-q_{2\perp}^2+i\epsilon}\sh{\bar{n}}\,
A^+(q_2-q_1)\sh{n}\,
\frac{iQ}{2Qq_1^+-q_{1\perp}^2+i\epsilon}\sh{\bar{n}}\,
A^+(q_1-q_0)\sh{n}\,
\xi_{\bar{n}}(q_0)
\,, 
\label{wilexp0}
\end{equation}
where $q_i$ are the collinear quark momenta after each scattering.
The Dirac spinor $\xi_{\bar{n}}(q_0)$ satisfies $\sh{\bar{n}}\,\xi_{\bar{n}}(q_0) =0$ 
and is normalized as $\xi^\dagger_{\bar{n}}(q_0)\,\xi_{\bar{n}}(q_0) = \sqrt{2}Q$.
As discussed in section~\ref{sec:pccov}, see eq.~(\ref{eqn:Gprop}) and
(\ref{eqn:sSource}), Glauber fields in momentum space are insensitive
to $q_i^+$ at lowest order in $\lambda$. Note that it is essential
for this observation to consider soft sources; for e.g. collinear
sources, the statement would not be true. On the other hand, no
further assumptions about the size of the medium are necessary in our
case.  Hence, the integration over $q_i^+$ yields
\begin{equation}
\int dy^+d^2y_\perp\prod_i dy_i^- \,\cdots\,
\theta(y^-_3-y^-_2)A^+(y^+,y_2^-,y_\perp)\, \theta(y^-_2-y^-_1)A^+(y^+,y_1^-,y_\perp)\xi_{\bar{n}}(q_0)
\,,
\label{wilexp}
\end{equation}
where the Glauber fields are now expressed in position space and the $\theta$ functions come from 
the  $i\epsilon$ prescription in the collinear quark propagators.
The same result also holds when considering the case of a 
collinear quark interacting at some point with a soft gluon. 
For illustration, suppose that this happens at the second gluon interaction; 
the corresponding diagram would then contain a term proportional to 
\begin{equation}
\int \prod_i \frac{d^4q_i}{(2\pi)^4} \,\cdots\, 
\frac{i}{2q_{2}^++i\epsilon}\sh{\bar{n}}\,
A^+(q_{2}-q_1)\sh{n}\,
\frac{iQ}{2Qq_1^+-q_{1\perp}^2+i\epsilon}\sh{\bar{n}}\,
A^+(q_1-q_{0}) \sh{n}\,\xi_{\bar{n}}(q_0)\,,
\label{wilexp1}
\end{equation}
where $A^+(q_{2}-q_1)$ is now a soft field, while $A^+(q_1-q_{0})$ is a Glauber field. 
For $q_{2}^+$ scales as a soft momentum and $q_{1}^+$ as a collinear one, we may neglect $q_{1}^+\sim Q\lambda^2$
with respect to $q_{2}^+\sim Q\lambda$ in the soft field. Hence, integrating over $q_i^+$ yields 
again eq. \eqref{wilexp}. 

Equation \eqref{wilexp} is just a term in the expansion of the Wilson line 
\begin{equation}
W[y^+,y_\perp]={\rm P}\,\exp\left[ig\int_{-L/\sqrt{2}}^{L/\sqrt{2}}\,dy^-A^+(y^+,y^-,y_\perp)\right]\,,
\label{eqn:longWilson}
\end{equation}
with the $\theta$ functions in \eqref{wilexp} ensuring the path ordering.
The distance $\sqrt{2}L$ is the distance that the parton travels along the light cone 
when moving through the medium in a box with a side length $L$. 

From eq. \eqref{Pkperp} it then follows that at leading order in the power counting, 
and in a covariant gauge, the probability $P(k_\perp)$ is given by 
\begin{equation}
P(k_\perp)=\int\,d^2x_\perp e^{ik_\perp\cdot x_\perp}\,
\frac{1}{N_c}
\left\langle \mathrm{Tr}\left\{W^\dagger[0,x_\perp]W[0,0]\right\}\right\rangle\,,
\label{eq:eramo}
\end{equation}
where $N_c=3$ is the number of colors. The expression holds for the propagation of 
a collinear quark, in which case the gluons in the Wilson lines are 
in the fundamental representation; the expression for the case of the propagation 
of a collinear gluon is similar and follows from replacing $1/N_c$ by $1/(N_c^2-1)$ 
and the gluons in the fundamental representation by gluons in the adjoint one.
The trace in \eqref{eq:eramo} has to be understood as a trace over color matrices.
Fields and color matrices are both path-ordered, which, in the plane $x^+=0$, 
also means that the fields in $W[0,0]$ are time ordered, while the fields in 
$W^\dagger[0,x_\perp]$ are anti-time ordered.

\begin{figure}[ht]
\center
\includegraphics[width=95mm]{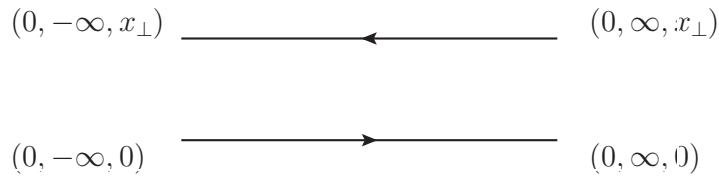}
\caption{The Wilson lines in the right-hand side of eq. (\ref{eq:eramo}):
the upper line corresponds to the operator $W^\dagger[0,x_\perp]$, the lower 
line to the operator $W[0,0]$, both taken in the limit $L\to\infty$.}
\label{fig:pcov}
\end{figure}

Although we agree with the result of \cite{D'Eramo:2010ak}, there are some differences worth emphasizing.
With respect to the derivation of the result, the step from \eqref{wilexp0} or  \eqref{wilexp1} to \eqref{wilexp} relies 
on the observation that Glauber gluons do not depend, at lowest order, on the momentum component 
along the light-cone direction $n$. In \cite{D'Eramo:2010ak}, instead, eq. \eqref{wilexp} follows from 
requiring $Q\,L \ll 1/\lambda^2$, a request that appears to us 
unnecessary when assuming soft sources.
Because in eq. \eqref{eqn:longWilson} we have chosen a symmetric integration region in $y^-$, 
taking $L\rightarrow\infty$ modifies $\displaystyle \int_{-L/\sqrt{2}}^{L/\sqrt{2}}\,dy^-$ into $\displaystyle \int_{-\infty}^{\infty}\,dy^-$.
Note that this limit is not allowed if $L$ is bound by an infrared cut-off.
In the limit $L\to\infty$, the Wilson lines in \eqref{eq:eramo} are shown in fig.~\ref{fig:pcov}.
With respect to the interpretation of the result, we note that if soft gluons contribute, their contribution 
would be encoded in the same average of Wilson lines that encodes the contribution from Glauber gluons.
This observation may be of relevance for perturbative calculations of \eqref{eq:eramo} beyond leading order. 

Finally, we mention that, while $W$ is formally similar to the usual collinear
Wilson line of SCET, $W_{\bar n}$, it has a different origin and the gluon field
$A^+$ is not collinear. Here, $W$ comes from resumming infinite Glauber gluons 
that interact with a collinear parton through a single vertex in the Lagrangian.
This leads to a Wilson line integrated along the same direction as the momentum of the initial high-energy parton. 
There, in contrast, the gauge invariant building block of SCET \cite{Hill:2002vw}, 
$W_{\bar n}^\dagger\xi_{\bar n}$, comes from resumming an infinite amount of $\bar n$-collinear gluons
attached to $n$-collinear quarks. This leads to a Wilson line integrated along the opposite light-cone direction.  

Equation \eqref{eq:eramo} was first derived in \cite{CasalderreySolana:2007zz,Liang:2008vz} within different approaches. 
Clearly the equation is not valid in light-cone gauge: if $A^+(x)=0$, 
the Wilson line  \eqref{eqn:longWilson} becomes equal to one and the jet quenching
parameter vanishes. This is not surprising since, as discussed in section \ref{sec:pclc}, 
new interaction terms show up at lowest order in that gauge and should be accounted for.

\subsection{Jet broadening in light-cone gauge}
\label{sec:lc}
In this section, we compute the broadening of the transverse momentum 
of a quark in the light-cone gauge $A^+(x)=0$. This will lead to 
a gauge-invariant generalization of eq.~(\ref{eq:eramo}).
We will organize the calculation in an expansion in powers of the background 
(Glauber) gluon fields in the medium, eventually resumming the expansion to all orders. 
An expansion in powers of the gluon fields is sometimes called opacity expansion. 
First, we compute the broadening at leading order in the opacity expansion, 
i.e. we consider the interaction of the collinear quark with one single Glauber gluon. 
Then we generalize the result to an arbitrary order.

\subsubsection{Leading order in the opacity expansion}
\begin{figure}[ht]
\center
\includegraphics[width=65mm]{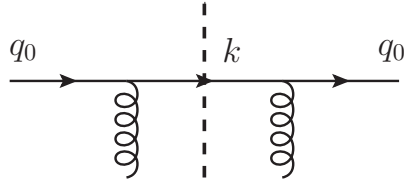}
\caption{Leading-order diagram in the opacity expansion.}
\label{fig:2g}
\end{figure}

We calculate the amplitude ${\cal A}_{11}$, i.e. the amplitude describing the propagation of a collinear quark in a medium 
of Glauber gluons at leading order in the opacity expansion.
The corresponding diagram is shown in fig.~\ref{fig:2g}.

Computing the diagram of fig.~\ref{fig:2g} is 
like computing the imaginary part of the forward scattering 
amplitude of a quark with momentum $q_0=(0,Q,0)$ on a background field. 
At the cut the quark acquires a transverse momentum $k_\perp$.  
Note that, for the calculation is performed in light-cone gauge, 
the quark couples to the transverse component of the gluon field 
and the quark-gluon vertex is at leading order given by eq.~\eqref{eq:feyn1}.
Gluons can be considered as background fields at this stage of the computation.

Following the definitions given at the beginning of section \ref{sec:jb}, \
the contribution of the diagram in fig.~\ref{fig:2g} to ${d^2{\cal A}_{11}}/{d^2k_\perp}$ reads
\begin{align}
\frac{d^2{\cal A}_{11}}{d^2k_\perp}=\frac{1}{\sqrt{2}QL^3}\int\frac{\,dk^+\,dk^-}{(2\pi)^2}&\;
2\pi\, Q\, \delta(2Qk^+-k_\perp^2)\;
\nonumber\\
&\hspace{-15mm}
\times\bar{\xi}_{\bar{n}}(q_0) \frac{gA_\perp^{\mathrm{sin}}(q_0-k)\cdot k_\perp}{2Q}\sh{n}
\; \sh{\bar{n}} \; \frac{k_\perp\cdot gA_\perp^{\mathrm{sin}}(k-q_0)}{2Q}\sh{n}\,\xi_{\bar{n}}(q_0)\,.
\end{align}
Simplifying the Dirac matrices and writing the gluon fields in position space we find
\begin{align}
\frac{d^2{\cal A}_{11}}{d^2k_\perp}=\frac{1}{2\sqrt{2}Q^3L^3}
\int\,d^4y\,d^4y'\;k_\perp^ik_\perp^j\, e^{iQ({y'_+}-y_+)}
\int\frac{\,dk^+\,dk^-}{(2\pi)^2}&\;2\pi\, Q\,\delta(2Qk^+-k_\perp^2)\,e^{ik\cdot(y-y')}
\nonumber\\ 
&\hspace{-58mm}
\times\bar{\xi}_{\bar{n}}(q_0)\,gA_\perp^{\mathrm{sin}\,i}({y^+}',{y^-}',{y_\perp}')
\,gA_\perp^{\mathrm{sin}\,j}(y^+,y^-,y_\perp)\sh{n}\,\xi_{\bar{n}}(q_0)\,.
\end{align}
The momentum integrals yield
\begin{equation}
\int\frac{\,dk^+\,dk^-}{(2\pi)^2}\;2\pi\, Q\,\delta(2Qk^+-k_\perp^2)\,e^{ikx}
=\frac{\delta(x^+)}{2}\;e^{i\big(\frac{k_\perp^2}{2Q}+i\,\mathrm{sgn}(x^-)\epsilon\big)x^-}e^{-ik_\perp\cdot x_\perp}\,,
\end{equation}
where the $i\epsilon$ prescription ensures that the function vanishes exponentially in the limit $|x^-|\to\infty$. 
Hence, we obtain 
\begin{align}
&\frac{d^2{\cal A}_{11}}{d^2k_\perp}=
\frac{1}{4\sqrt{2}Q^3L^3}\int\,dy^+\,d^2y_\perp\,d^2{y'_\perp}\,e^{-ik_\perp\cdot(y_\perp-{y'_\perp})}\,k_\perp^ik_\perp^j
\nonumber\\
&\hspace{20mm}
\times\bar{\xi}_{\bar{n}}(q_0)\left[\int_{-\infty}^\infty\,d{y^-}'\,gA_\perp^{\mathrm{sin}\,i}(y^+,{y^-}',{y_\perp}')
\,e^{-i\frac{k_\perp^2 }{2Q}{y^-}'}\right]
\nonumber\\
&\hspace{31mm}
\times\left[\int_{-\infty}^\infty\,d{y^-}\,gA_\perp^{\mathrm{sin}\,j}(y^+,{y^-},{y_\perp})
\,e^{i\frac{k_\perp^2 }{2Q}{y^-}}\right]e^{-\epsilon|y^--{y^-}'|}\,
\sh{n}\,\xi_{\bar{n}}(q_0)\,.
\end{align}
Using eqs. \eqref{eqn:gluonDecomp} and \eqref{eqn:puregauge}, we may perform 
the integration over $y^-$ and ${y^-}'$ obtaining 
\begin{align}
\frac{d^2{\cal A}_{11}}{d^2k_\perp}=\frac{1}{\sqrt{2}QL^3}\int\,dy^+\,d^2y_\perp\,d^2{y'_\perp}
&\,e^{-ik_\perp\cdot(y_\perp-{y'_\perp})}
\nonumber\\
&\hspace{-150pt}
\times\bar{\xi}_{\bar{n}}(q_0)\left[g\phi^+(y^+,{y'_\perp})-g\phi^-(y^+,{y'_\perp})\right]
\left[g\phi^+(y^+,y_\perp)-g\phi^-(y^+,y_\perp)\right]\sh{n}\,\xi_{\bar{n}}(q_0)\,.
\end{align}

\subsubsection{Arbitrary order in the opacity expansion}
\begin{figure}[ht]
\center
\includegraphics[width=150mm]{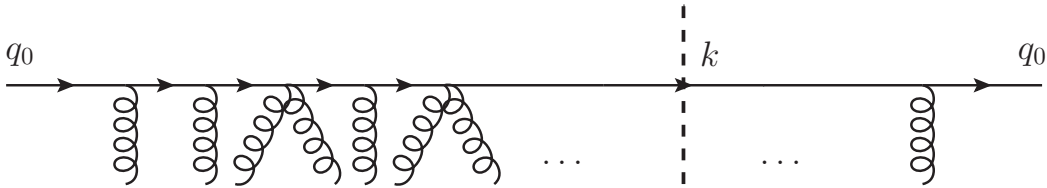}
\caption{Arbitrary order diagram in the opacity expansion in light-cone gauge. 
Any number of 1- or 2-gluon vertices can appear on both sides of the cut.}
\label{fig:full}
\end{figure}

The main difficulty in generalizing the previous result to an
arbitrary order comes from the fact that the interaction between
collinear quarks and Glauber gluons can be mediated in light-cone gauge 
by 1-gluon or 2-gluon vertices. The Feynman rule for the leading-order 1-gluon vertex 
is given in eq.~(\ref{eq:feyn1}), whereas that one for the leading-order 2-gluon vertex 
is in \eqref{eq:feyn2}. The most general diagram contributing to 
$d^2{\cal A}_{nm}/d^2k_\perp$ is of the type shown in fig.~\ref{fig:full}.

Let us first consider Feynman diagrams with $n$ gluons on the left of the cut 
(i.e. the part with the ingoing on-shell quark). 
These are all the Feynman diagrams made of an incoming collinear quark with momentum $(0,Q,0)$, 
$n$ gluons attached to the quark line, and an outgoing collinear quark with momentum
$({k_\perp^2}/{(2Q)},k^-,k_\perp)$, $k^-$ being the sum of $Q$ and a residual momentum of order $Q\lambda$
or smaller. If $n=1$, there is only one possible Feynman diagram, the one shown in fig.~\ref{fig:2g} and 
computed in the previous section. If $n=2$, there are two possible diagrams:
one made of two 1-gluon vertices and one made of a 2-gluon vertex.
If $n=3$, the possible diagrams are three, if $n=4$ the possible diagrams are five and so on.
In fact, the number of diagrams with $n$ gluons attached to the quark line 
is $F_{n+1}$, where $F_n$ is the Fibonacci number, implying that the number of diagrams 
grows exponentially for large $n$.

An important observation is that, due to the $i\epsilon$ prescription of the quark propagator, 
the fields on the left of the cut are time ordered. This means that fields evaluated at
$y^-=-\infty$ (the $\phi^-$ fields) are on the right and fields evaluated at $y^-=\infty$
(the $\phi^+$ fields) are on the left in the expression of the amplitude.
We call this amplitude $G_n(k^-,k_\perp)$, i.e. the sum of all diagrams with $n$ gluons attached 
to the quark line on the left of the cut. Hence, $G_n(k^-,k_\perp)$ can be written as 
\begin{equation}
G_n(k^-,k_\perp)= 
\sum_{j=0}^n\int\frac{\,d^4q}{(2\pi)^4}\,G^+_{n-j}(k^-,k_\perp ,q)\,\frac{iQ}{2Qq^+-q_\perp^2+i\epsilon}\,\sh{\bar{n}}\,G^-_j(q)\,,
\label{eq:sum}
\end{equation}
where $G^+_n$ collects all insertions of $\phi^+$ fields and $G^-_n$ all insertions of $\phi^-$ fields.
Once convoluted with the cut and the amplitude on the right of the cut,  $G_n(k_\perp)$ provides ${d^2{\cal A}_{nm}}/{d^2k_\perp}$.

First, we calculate $G^-_n$; it fulfills the recursion relation 
\begin{align}
G^-_n(q)=&\int\frac{\,d^4q'}{(2\pi)^4}\,G^-_{n-1}(q')\times\raisebox{-30pt}{\includegraphics[width=45mm]{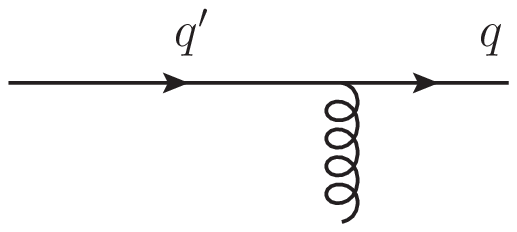}}
\nonumber\\
+&\int\frac{\,d^4q''}{(2\pi)^4}\,G^-_{n-2}(q'')\times\raisebox{-30pt}{\includegraphics[width=45mm]{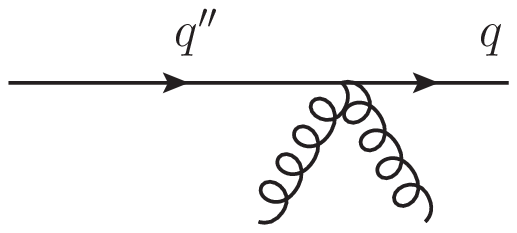}},
\label{eq:rel-}
\end{align}
where the Feynman graphs have to be understood as the product of a (momentum space) propagator (on the left)
and a 1-gluon or a 2-gluon vertex. The solution of the equation is
\begin{equation}
G^-_n(q)=\int dy^+\,dy^-\,\theta(-y^-)e^{i(q^--Q)y^++iq^+y^-}f_n(y^+,q_\perp)\sh{n}\,,
\label{eq:res-}
\end{equation}
with
\begin{equation}
f_n(y^+,q_\perp)=\frac{i}{2Qn!}\,q_\perp^2\int d^2y_\perp\, e^{-iq_\perp\cdot y_\perp}
\,{\rm P}\left(\left[ig\phi^-(y^+,y_\perp)\right]^n\right)\,.
\label{eq:res1-}
\end{equation}
The path-ordering P refers to the fields $A_\perp$ that appear in the definition of $\phi$ (see eq.~(\ref{eqn:phi})):
\begin{eqnarray}
{\rm P}\left(\left[\phi^\pm(x^+,x_\perp)\right]^n\right) &=& 
(-1)^n\,n!\, \int_{-\infty}^0 ds_1\, \dots \int_{-\infty}^{s_{n-1}} ds_n\, 
l_\perp \cdot A_\perp(x^+,\pm\infty,x_\perp+l_\perp s_1)
\nonumber\\
&&\hspace{37mm}
\times \dots l_\perp\cdot A_\perp(x^+,\pm\infty,x_\perp+l_\perp s_n)
\,.
\end{eqnarray}
In particular, it holds that 
\begin{equation}
\nabla^i_\perp{\rm P} \left(\left[\phi^\pm(x^+,x_\perp)\right]^n\right)
= n \left(\nabla^i_\perp \phi^\pm(x^+,x_\perp)\right) {\rm P} \left(\left[\phi^\pm(x^+,x_\perp)\right]^{n-1}\right)\,.
\label{derPphi}
\end{equation}
The proof of eqs. \eqref{eq:res-} and \eqref{eq:res1-} is given in Appendix \ref{sec:ape-}. 

The function $G^+_n$ can be calculated in a similar manner. It satisfies the recursion relation 
\begin{align}
G^+_n(k^-,k_\perp ,q)=&\int\frac{\,d^4q'}{(2\pi)^4}
\raisebox{-27pt}{\includegraphics[width=37mm]{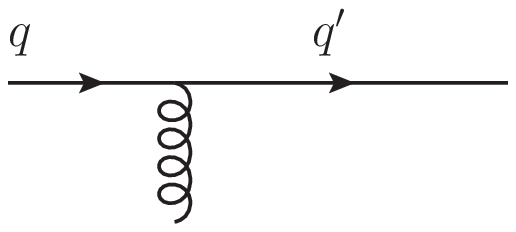}}
\times G^+_{n-1}(k^-,k_\perp , q')
\nonumber\\
+&\int\frac{\,d^4q''}{(2\pi)^4}\raisebox{-27pt}{\includegraphics[width=37mm]{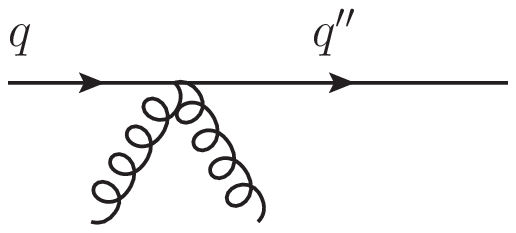}}
\times G^+_{n-2}(k^-,k_\perp ,q'')
\,,
\end{align}
where the Feynman graphs have to be understood now as the product of a 
1-gluon or a 2-gluon vertex and a (momentum space) propagator (on the right). 
The solution of the equation is
\begin{equation}
G^+_n(k^-,k_\perp ,q)=\int dy^+\,dy^-e^{i(k^--q^-)y^++i\left(\frac{k_\perp^2}{2Q}-q^+\right)y^-}\theta(y^-)\,g_n(y^+,q_\perp,k_\perp)\sh{n}\,,
\label{eq:res+}
\end{equation}
with
\begin{equation}
g_n(y^+,q_\perp,k_\perp)=\frac{-i}{2Qn!}\,(k_\perp^2-q_\perp^2)\int d^2y_\perp
\, e^{-i(k_\perp-q_\perp)\cdot y_\perp}\bar{\rm P}\left(\left[-ig\phi^+(y^+,y_\perp)\right]^n\right)\,.
\label{eq:res1+}
\end{equation}
The symbol $\bar{\rm P}$ stands for the anti-path ordering operator and refers to the fields $A_\perp$ 
that appear in the definition of $\phi$:
\begin{eqnarray}
\bar{\rm P}\left(\left[\phi^\pm(x^+,x_\perp)\right]^n\right) &=& 
(-1)^n\,n!\, \int_{-\infty}^0 ds_1\, \dots \int_{-\infty}^{s_{n-1}} ds_n\, 
l_\perp \cdot A_\perp(x^+,\pm\infty,x_\perp+l_\perp s_n)
\nonumber\\
&&\hspace{37mm}
\times \dots l_\perp\cdot A_\perp(x^+,\pm\infty,x_\perp+l_\perp s_1)
\,.
\end{eqnarray}
In this case, it holds that 
\begin{equation}
\nabla^i_\perp\bar{\rm P} \left(\left[\phi^\pm(x^+,x_\perp)\right]^n\right)
= n \,\bar{\rm P} \left(\left[\phi^\pm(x^+,x_\perp)\right]^{n-1}\right)\left(\nabla^i_\perp \phi^\pm(x^+,x_\perp)\right)\,.
\end{equation}

Substituting the obtained expressions for $G^-_n$ and $G^+_n$ into eq. (\ref{eq:sum}), we obtain 
\begin{align}
G_n(k^-,k_\perp) &= \int\,dy^+\,d^2y_\perp\,e^{i(k^--Q)y^+}e^{-i k_\perp\cdot y_\perp}
\nonumber\\
&\times \sum_{j=0}^n \,\frac{1}{j!(n-j)!}\,
\bar{\rm P}\left( \big[-ig\phi^+(y^+,y_\perp)\big]^j\right)
\,{\rm P}\left( \big[ig\phi^-(y^+,y_\perp)\big]^{n-j}\right)\sh{n}\,.
\label{Gneq}
\end{align}
This is the amplitude for $n$ gluons attached to the quark on the left of the cut. 
The amplitude for $m$ gluons attached to the quark
on the right of the cut just follows from the Hermitian conjugate of \eqref{Gneq}. 
Convoluting the two amplitudes and the cut over all intermediate momenta, 
we get $d^2{\cal A}_{nm}/d^2k_\perp$:
\begin{eqnarray}
\frac{d^2{\cal A}_{mn}}{d^2k_\perp}&=&
\frac{1}{\sqrt{2}QL^3}\int\frac{\,dk^+\,dk^-}{(2\pi)^2}\; 2\pi\, Q\, \delta(2Qk^+-k_\perp^2)
\; \bar{\xi}_{\bar{n}}(q_0)
G_m^\dagger(k^-,k_\perp)\,\sh{\bar{n}}\,G_n(k^-,k_\perp)
{\xi}_{\bar{n}}(q_0)
\nonumber\\
&=& \frac{1}{\sqrt{2}QL^3}\int\,dy^+\,d^2y'_\perp\,d^2y_\perp\, e^{-ik_\perp\cdot(y_\perp-y'_\perp)}
\nonumber\\
&& \hspace{-0.3cm}
\times \bar{\xi}_{\bar{n}}(q_0) \sum_{j'=0}^m\,\frac{1}{(m-j')!j'!}\,
\bar{\rm P}\left( \big[-ig\phi^-(y^+,y'_\perp)\big]^{m-j'}\right)
\,{\rm P}\left( \big[ig\phi^+(y^+,y'_\perp)\big]^{j'}\right)
\nonumber\\
&& \hspace{-0.3cm}
\times \sum_{j=0}^n\,\frac{1}{j!(n-j)!}\,
\bar{\rm P}\left( \big[-ig\phi^+(y^+,y_\perp)\big]^{j}\right)
\,{\rm P}\left( \big[ig\phi^-(y^+,y_\perp)\big]^{n-j}\right)
\sh{n}\,\xi_{\bar{n}}(q_0)\,.
\end{eqnarray}

\begin{figure}[ht]
\center
\includegraphics[width=120mm]{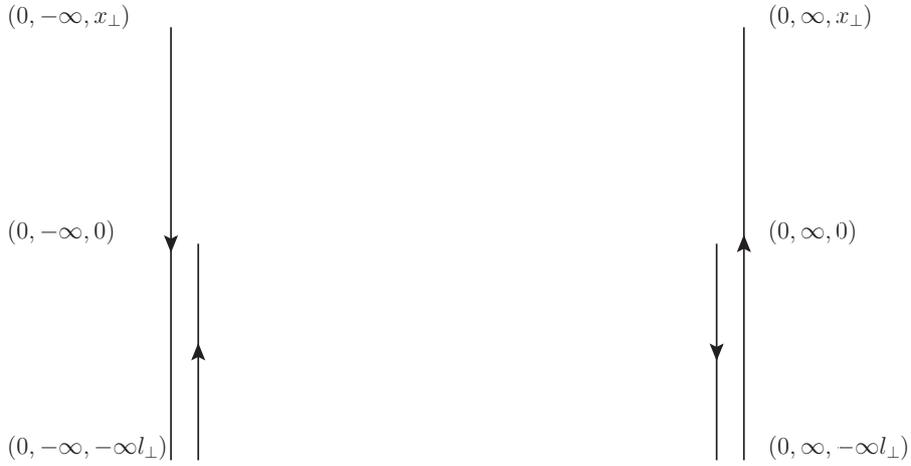}
\caption{Transverse Wilson lines contributing to the right-hand side of eq. (\ref{eq:Plc}).
For convenience, in the picture, we have chosen $l_\perp \parallel x_\perp$.
}
\label{fig:lc}
\end{figure}

From  $d^2{\cal A}_{nm}/d^2k_\perp$ and eq. \eqref{Pkperp}, it follows that the transverse 
momentum broadening probability, $P(k_\perp)$, is given by 
\begin{align}
P(k_\perp)=
\int\,d^2x_\perp&\, e^{ik_\perp\cdot x_\perp}
\nonumber\\
&\hspace{-2.3cm}
\times\frac{1}{N_c}\left\langle\mathrm{Tr}\left\{
\bar{\rm P}\left( e^{-ig\phi^-(0,x_\perp)}\right) {\rm P}\left( e^{ig\phi^+(0,x_\perp)}\right)
\bar{\rm P}\left( e^{-ig\phi^+(0,0)}\right) {\rm P}\left( e^{ig\phi^-(0,0)}\right)
\right\}\right\rangle\,,
\label{eq:plight}
\end{align}
where the trace refers to the color matrices and follows from averaging over the colors 
of the initial state. We have also made use of 
\begin{equation}
\bar{\rm P}\left(e^{-ig\phi^+}\right){\rm P}\left(e^{ig\phi^-}\right)
=\sum_{n=0}^\infty\,\sum_{j=0}^n\,\frac{1}{j!(n-j)!}\,\bar{\rm P}\left([-ig\phi^+]^j\right)
\,{\rm P}\left([ig\phi^-]^{n-j}\right).
\end{equation}
By means of eq. \eqref{eqn:phi} we can express  $P(k_\perp)$ in terms of the gluon fields $A_\perp$ at light-cone infinity: 
\begin{align}
P(k_\perp)=\int\,d^2x_\perp\, e^{ik_\perp\cdot x_\perp}
\frac{1}{N_c}\left\langle\mathrm{Tr}\left\{ 
T^\dagger(0,-\infty,x_\perp)\,T(0,\infty,x_\perp)\,T^\dagger(0,\infty,0)\,T(0,-\infty,0) \right\}\right\rangle\,,
\label{eq:Plc}
\end{align}
where $T$ is the transverse Wilson line \cite{Idilbi:2010im}
\begin{equation}
T(x^+,\pm\infty,x_\perp) = {\rm P}\,\exp\left[-ig\int_{-\infty}^0 ds\; l_\perp\cdot A_\perp(x^+,\pm\infty,x_\perp+l_\perp s)\right]\,.
\label{eqn:TWilson}
\end{equation}
Note that color matrices and operators are path ordered.
The Wilson lines in \eqref{eq:Plc} are shown in fig.~\ref{fig:lc}.

\subsection{Jet broadening in arbitrary gauge}
\label{sec:jball}
In the last section, we have derived in light-cone gauge the probability for a collinear quark 
to gain a certain transverse momentum while travelling through a medium. 
For the general gauge case, one has also to include the interaction between 
the collinear quark and Glauber and soft fields $A^+$, which is encoded 
in the first term of the SCET Lagrangian \eqref{eq:SCET}. 
This is done by appropriately extending eq.~\eqref{eq:sum} and noticing that 
in fig.~\ref{fig:Amn}  the operators on the left of the cut 
are all time ordered while the operators on the right of the cut are all anti-time ordered. 
As a consequence,  on the left of the cut, the transverse contributions to $G^+_n$ always appear to the left 
and the transverse contributions to $G^-_n$ always appear to the right of the interactions with the $A^+$ fields.
This allows combining the result obtained in section~\ref{sec:Pcg} in covariant gauge 
with the result obtained in section~\ref{sec:lc} in light-cone gauge 
to write a fully gauge-invariant expression for $P(k_\perp)$.
In the $L\to\infty$ limit, it reads
\begin{align}
P(k_\perp)=&\int\,d^2x_\perp\, e^{ik_\perp\cdot x_\perp}
\nonumber\\
&\hspace{-5mm}
\times 
\frac{1}{N_c}\left\langle\mathrm{Tr}\left\{
T^\dagger(0,-\infty,x_\perp)\, W^\dagger[0,x_\perp]\, T(0,\infty,x_\perp)\;
T^\dagger(0,\infty,0)\, W[0,0]\, T(0,-\infty,0) \right\}\right\rangle
\nonumber\\
=&\int\,d^2x_\perp\, 
e^{ik_\perp\cdot x_\perp}
\frac{1}{N_c}\left\langle\mathrm{Tr}\left\{ {\mathcal W}^\dagger(x_\perp)\;{\mathcal W}(0) \right\}\right\rangle\,,
\label{eqn:complete}
\end{align}
where $W$ has been given in \eqref{eqn:longWilson},  $T$ in  \eqref{eqn:TWilson}, 
and we have defined ${\mathcal W}(x_\perp)$  $=$ $T^\dagger(0,\infty,x_\perp)\,$ $\times W[0,x_\perp]\,$ $T(0,-\infty,x_\perp)\,.$

\begin{figure}[ht]
\center
\includegraphics[width=120mm]{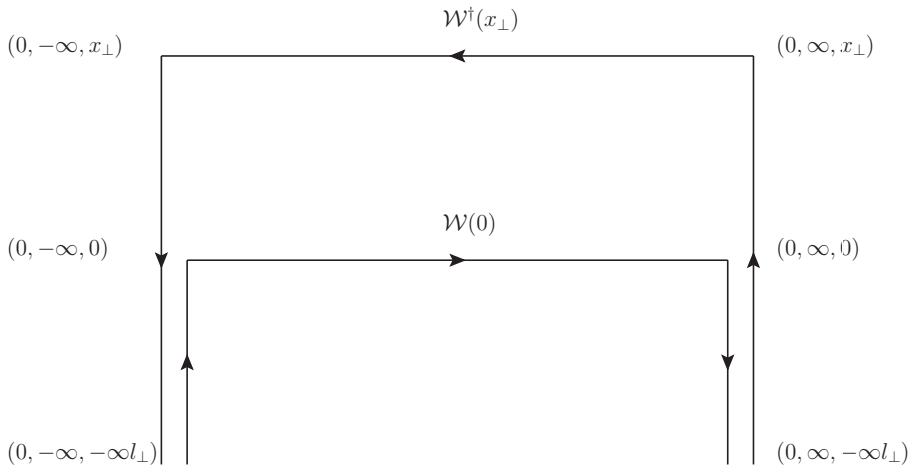}
\caption{Wilson lines appearing in eq. \eqref{eqn:complete}. 
Horizontal lines are oriented along the $\bar{n}$ direction: 
they correspond to the Wilson lines defined in \eqref{eqn:longWilson}.
Vertical lines extend in the transverse plane along the direction of the vector $l_\perp$:
they correspond to the Wilson lines defined in \eqref{eqn:TWilson}.
For convenience, in the picture, we have chosen $l_\perp \parallel x_\perp$.
}
\label{fig:p2}
\end{figure}

The Wilson lines appearing in eq.~\eqref{eqn:complete} are shown in fig.~\ref{fig:p2}.
Operators and color matrices are path-ordered along the lines.\footnote{
It has been remarked in \cite{D'Eramo:2010ak} that, in the case of equilibrium 
thermal field averages, a way to handle the 
fact that the fields in the Wilson lines of $P(k_\perp)$ are not time ordered is 
to express the thermal average in the so-called real-time formalism (see e.g. \cite{LeBellac:1996,Landsman:1986uw,Thoma:2000dc}).
This amounts to modifying the integration path along the imaginary-time axis in the 
partition function to include the real-time axis. More specifically, the modification adds to the 
imaginary-time path a path along the real-time axis at zero imaginary time and a parallel path, 
oppositely oriented, at imaginary time $-i\epsilon$.
In our case, the correct ordering of eq.~\eqref{eqn:complete} would be ensured by 
locating the fields of ${\mathcal W}(0)$ on the zero imaginary-time real-axis, 
and the fields of ${\mathcal W}^\dagger(x_\perp)$ on the  $-i\epsilon$ imaginary-time axis.
Fields located at different imaginary times can be treated independently.
This procedure can be extended to out of equilibrium situations (see e.g. \cite{Landsman:1986uw,Berges:2004yj}).
}
In the case of non-singular gauges, the transverse fields vanish at $L\to\infty$, 
the transverse Wilson lines become one, and the expression of $P(k_\perp)$ reduces to 
eq. \eqref{eq:eramo}. In the case of the light-cone gauge, $A^+(x)=0$, the Wilson lines 
along the light-cone direction $\bar{n}$ become one, and the expression of $P(k_\perp)$ reduces to 
eq. \eqref{eq:Plc}.

\begin{figure}[ht]
\center
\includegraphics[width=130mm]{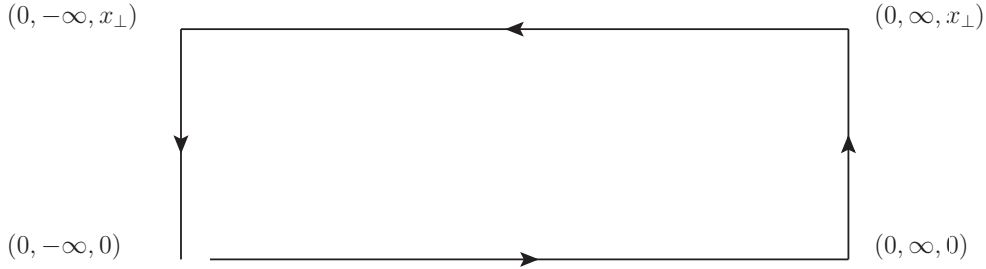}
\caption{Wilson lines contributing to the right-hand side of eq.~\eqref{eqn:complete3}.
Longitudinal lines are like in fig.~\ref{fig:p2}; vertical lines stand for the 
Wilson lines defined in \eqref{vertxy}.}
\label{fig:p3}
\end{figure}

It is possible to arrange the Wilson lines in eq.~\eqref{eqn:complete} in several equivalent ways. 
First, we recall that, according to \eqref{eqn:phi} and the following 
discussion, the transverse Wilson lines are independent of the path chosen to connect the initial 
and the final point. This freedom allows us to deform the contour of $T(0,\infty,x_\perp)$ such that 
\begin{equation}
T(0,\infty,x_\perp) = [x_\perp,0_\perp]_+ \; T(0,\infty,0) \;  [-\infty l_\perp,x_\perp - \infty l_\perp]_+ \,,
\end{equation}
where we have defined 
\begin{equation}
[x_\perp,y_\perp]_\pm = {\rm P}\,\exp\left[-ig\int_{1}^0 ds\; (y_\perp-x_\perp)\cdot A_\perp(0,\pm\infty,x_\perp+(y_\perp-x_\perp) s)\right]\,.
\label{vertxy}
\end{equation}
Because fields at infinite distance in the transverse plane vanish also in light-cone gauge, it holds that 
$[-\infty l_\perp,x_\perp - \infty l_\perp]_+ =1$ and we can write 
\begin{eqnarray}
P(k_\perp) &=& \int\,d^2x_\perp\, e^{ik_\perp\cdot x_\perp}
\nonumber\\
&& \times \frac{1}{N_c}
\left\langle\mathrm{Tr}\left\{
T^\dagger(0,-\infty,x_\perp)\, W^\dagger[0,x_\perp]\, [x_\perp,0]_+\, W[0,0]\, T(0,-\infty,0) \right\}\right\rangle\,.
\label{eqn:complete2}
\end{eqnarray}
Also the transverse Wilson lines at the very left and right of (\ref{eqn:complete2}) 
combine in a similar way, but because the trace only refers to the color matrices and not 
to the field operators, the argument requires some care. It goes as follows.
First, we deform the contour of $T^\dagger(0,-\infty,x_\perp)$ into ($[x_\perp,0]^\dagger_-=[0,x_\perp]_-$)
\begin{equation}
T^\dagger(0,-\infty,x_\perp) = T^\dagger(0,-\infty,0)[0,x_\perp]_-\,,
\end{equation}
then we rewrite the trace in \eqref{eqn:complete2} as
\begin{equation}
\mathrm{Tr}\left\{ 
T^\dagger(0,-\infty,0)\,[0,x_\perp]_-\, W^\dagger[0,x_\perp]\, [x_\perp,0]_+\, W[0,0]\, T(0,-\infty,0) 
\right\}\,.
\label{eqn:trace}
\end{equation}
The $N$-th term in the expansion of the two transverse Wilson lines in (\ref{eqn:trace}) is 
\begin{eqnarray}
&& \sum_{n=0}^N \mathrm{Tr}\Bigg\{
(ig)^n  \int_{-\infty}^0 ds_n\int_{s_n}^0 ds_{n-1}\dots\int_{s_2}^0 ds_{1} 
\nonumber\\
&& \hspace{10mm}
\times l_\perp\cdot A^{a_n}_\perp(0,-\infty,l_\perp s_n)T^{a_n}\dots l_\perp\cdot A^{a_1}_\perp(0,-\infty,l_\perp s_1)T^{a_1}\;[\dots]
\nonumber\\
&& \hspace{10mm}
\times (-ig)^{N-n} \int_{-\infty}^0 ds_{N}\int_{-\infty}^{s_N} ds_{N-1}\dots\int_{-\infty}^{s_{n+2}} ds_{n+1}
\nonumber\\
&& \hspace{10mm}
\times l_\perp\cdot A^{a_N}_\perp(0,-\infty,l_\perp s_N)T^{a_N}\dots l_\perp\cdot A^{a_{n+1}}_\perp(0,-\infty,l_\perp s_{n+1})T^{a_{n+1}}
\Bigg\}\,,
\label{eqn:trace2}
\end{eqnarray}
where the $[\dots]$ stands for everything that is in between the transverse Wilson lines; 
the limits of integration ensure the proper path ordering. 
It is now crucial to note that $[\dots]$ contains only gauge fields whose 
coordinates project in the transverse plane on the straight line connecting 
$0$ to $x_\perp$, while all gauge fields from the transverse Wilson lines are evaluated at points that 
project outside of that line. Moreover, all fields are
evaluated at $x^+=0$. These two observations guarantee that the separation between 
the fields in $[\dots]$ and the fields from the transverse Wilson lines 
is space-like. Hence, the gauge fields from the transverse Wilson lines 
commute with $[\dots]$. By also using the cyclicity of the trace we can then rewrite (\ref{eqn:trace2}) as
\begin{align}
(ig)^N & \sum_{n=0}^N (-1)^{N-n}\, 
\int_{-\infty}^0 ds_{N}\int_{-\infty}^{s_N} ds_{N-1}\dots\int_{-\infty}^{s_{n+2}} ds_{n+1}
\int_{-\infty}^0 ds_n\int_{s_n}^0 ds_{n-1}\dots\int_{s_2}^0 ds_{1}
\nonumber\\
&\hspace{30mm}
\times l_\perp\cdot A^{a_{n}}(0,-\infty,l_\perp s_{n})\dots l_\perp\cdot A^{a_{1}}(0,-\infty,l_\perp s_{1})
\nonumber\\
&\hspace{30mm}
\times l_\perp\cdot A^{a_{N}}(0,-\infty,l_\perp s_{N})\dots l_\perp\cdot A^{a_{n+1}}(0,-\infty,l_\perp s_{n+1})
\nonumber\\
&\hspace{30mm}
\times \mathrm{Tr}\left\{[\dots]\,T^{a_{N}}\dots T^{a_{n+1}} T^{a_{n}}\dots T^{a_{1}} \right\}\,.
\end{align}
Since all gauge fields in the second and third line are each separated by a space-like interval,
they commute as well, and we find
\begin{align}
(ig)^N\sum_{n=0}^N (-1)^{N-n}\,&
\int_{-\infty}^0 ds_{N}\int_{-\infty}^{s_N} ds_{N-1}\dots\int_{-\infty}^{s_{n+2}} ds_{n+1}
\int_{-\infty}^0 ds_n\int_{s_n}^0 ds_{n-1}\dots\int_{s_2}^0 ds_{1}
\nonumber\\
&\hspace{30mm}
\times l_\perp\cdot A^{a_{1}}(0,-\infty,l_\perp s_{1})\dots l_\perp\cdot A^{a_{N}}(0,-\infty,l_\perp s_{N})
\nonumber\\
& \hspace{30mm}
\times  \mathrm{Tr}\left\{[\dots]\, T^{a_{N}}\dots T^{a_{1}} \right\}\,,
\end{align}
where only the first line depends on the summation index $n$. The sum adds up to zero.
This implies that only the zeroth-order term in the expansion of the transverse Wilson lines contributes and that 
$T^\dagger(0,-\infty,0)$ cancels with $T(0,-\infty,0)$ in (\ref{eqn:trace}). 
Equation \eqref{eqn:complete} can thus be rewritten in the equivalent way 
\begin{equation}
P(k_\perp)=\int\,d^2x_\perp\, e^{ik_\perp\cdot x_\perp}
\frac{1}{N_c}
\left\langle\mathrm{Tr}\left\{[0,x_\perp]_-\, W^\dagger[0,x_\perp]\, [x_\perp,0]_+\, W[0,0] \right\}\right\rangle\,.
\label{eqn:complete3}
\end{equation}
The advantage of eq. \eqref{eqn:complete3} is that it is explicitly independent of the choice for the vector $l_\perp$
and of the starting point of the integration path in \eqref{eqn:phi}.
The Wilson lines contributing to the right-hand side of \eqref{eqn:complete3} are shown in fig.~\ref{fig:p3}.
Note that, while the fields evaluated at $(0,\infty,0)$ in $[x_\perp,0]_+$ and $W[0,0]$ are contiguous, 
this is not the case for the fields evaluated at $(0,-\infty,0)$ in $[0,x_\perp]_-$ and $W[0,0]$, 
for they are separated from each other by fields located at light-like distance.
In fig.~\ref{fig:p3}, this is signaled by the small gap at $(0,-\infty,0)$. 

We can now prove the gauge invariance of $P(k_\perp)$. A convenient expression to use is 
eq. \eqref{eqn:complete2}. Under a gauge transformation $\Omega(x)$, $P(k_\perp)$ transforms as 
\begin{eqnarray}
P(k_\perp) &\to& \int\,d^2x_\perp\, e^{ik_\perp\cdot x_\perp}
\frac{1}{N_c}\left\langle\mathrm{Tr}\left\{
\Omega(0,-\infty,-\infty l_\perp)
T^\dagger(0,-\infty,x_\perp)\, W^\dagger[0,x_\perp]\, \right.\right.
\nonumber\\
&&
\hspace{25mm}
\left.\left.
\times [x_\perp,0]_+\, W[0,0]\, T(0,-\infty,0) \Omega^\dagger(0,-\infty,-\infty l_\perp)
\right\}\right\rangle\,,
\end{eqnarray}
where we have put to zero fields evaluated at infinite distance in the transverse plane.
The cancellation of the gauge transformations to the very left and right 
follows from noti\-cing that they are evaluated at $(0,-\infty,-\infty l_\perp)$
and, therefore, commute with all the fields in the Wilson lines, since they are separated by space-like intervals.
The conclusion is that $P(k_\perp)$ defined via eqs. \eqref{eqn:complete}, \eqref{eqn:complete2}
or \eqref{eqn:complete3} is gauge invariant.

Finally, we observe that, although the obtained gauge invariant expression of $P(k_\perp)$ reflects 
expectations on the gauge invariant completion of eq. (\ref{eq:eramo}) 
with transverse Wilson lines at $x^-=\pm\infty$ (in the large $L$ limit), 
the result cannot be found in this form in the literature. 
For instance, an expression similar to  \eqref{eqn:complete} was found in~\cite{Liang:2008vz}
by extending an analogous study on SIDIS to jet quenching.  
The expression in~\cite{Liang:2008vz} contains, however, 
only one transverse Wilson line at $x^-=-\infty$ but none at
$x^-=\infty$.  The difference with our result might be traced back to the
regularization of the singularity in light-cone gauge. When choosing
an asymmetric prescription, the perpendicular component of the gauge
field can vanish at either $+\infty^-$ or $-\infty^-$. In such a case, 
one of the transverse Wilson lines becomes one.

\subsection{\texorpdfstring{Jet quenching parameter $\hat{q}$}{}}
We show now how the obtained expression for $P(k_\perp)$ translates into the jet quenching parameter $\hat{q}$. 
According to \eqref{eq:qhat} and \eqref{eqn:complete3}, $\hat{q}$ can be written as
\begin{eqnarray}
\hat{q}&=&
\frac{1}{L^3}\int\frac{d^2k_\perp}{(2\pi)^2}\,d^2x_\perp \,d^2y_\perp e^{ik_\perp\cdot (x_\perp-y_\perp)}
\nonumber
\\
&& \hspace{10mm}
\times\frac{1}{N_c}\nabla_{x_\perp}\nabla_{y_\perp}
\left\langle\mathrm{Tr}\left\{[y_\perp,x_\perp]_-\, W^\dagger[0,x_\perp]\, [x_\perp,y_\perp]_+\, W[0,y_\perp] \right\}\right\rangle
\,.
\label{eq:newqhat}
\end{eqnarray}
By explicitly evaluating the derivatives acting on the Wilson lines,  
we find that Glauber and soft gluons contribute to the propagation of a collinear quark in a medium 
along a length $L\to\infty$ by 
\begin{eqnarray}
\hat{q}&=& \sqrt{2}\int\frac{d^2k_\perp}{(2\pi)^2}\,d^2x_\perp\,dx^- e^{i k_\perp\cdot x_\perp}
\nonumber\\
&&\hspace{5mm}
\times\frac{1}{N_c}\Bigg\langle \mathrm{Tr}\Bigg\{
[0,x_\perp]_-U^\dagger_{x_\perp}[x^-,-\infty]\,gF_\perp^{+i}(0,x^-,x_\perp)\,U^\dagger_{x_\perp}[\infty,x^-]
\nonumber\\
&&\hspace{22mm}
\times [x_\perp,0]_+U_{0_\perp}[\infty,0]\,gF_\perp^{+i}(0,0,0)\,U_{0_\perp}[0,-\infty]
\Bigg\}\Bigg\rangle\,,
\label{eq:FFkperp}
\end{eqnarray}
where $F_\perp^{+i}= \bar{n}\cdot\partial A_\perp^i-\nabla_\perp^iA^+ + ig[A^+,A_\perp^i]$. 
We have also defined the Wilson line 
\begin{equation}
U_{x_\perp}[x^-,y^-] = {\rm P}\,\exp\left[ig\int_{y^-}^{x^-}dz^- \; A^+(0,z^-,x_\perp)\right]\,,
\end{equation}
which is such that $U_{x_\perp}^\dagger[x^-,y^-] = U_{x_\perp}[y^-,x^-]$ and $U_{x_\perp}[L/\sqrt{2},-L/\sqrt{2}] = W[0,x_\perp]$.
Equation \eqref{eq:FFkperp} agrees at leading order in the opacity 
expansion with a similar expression that can be found in \cite{Idilbi:2008vm}.

The integral over $k_\perp$ in (\ref{eq:FFkperp}) has an ultraviolet 
cut-off, $q^{\rm max}$, which is of the order of $Q\lambda$, the size of the transverse 
momentum broadening that we are considering.
If this cut-off could be set to infinity, which may happen in dimensional regularization 
if the integral involves only transverse-momentum regions of order $Q\lambda$ or smaller, 
then the integral in $k_\perp$ leads to a delta function in the transverse coordinate 
that squeezes the contour of fig.~\ref{fig:p3} on one line. Under this condition 
$\hat{q}$ can be written as 
\begin{eqnarray}
\hat{q}&=& \sqrt{2}\int dx^- \frac{1}{N_c}\left\langle \mathrm{Tr}\left\{
U_{0_\perp}[-\infty,x^-]\,gF_\perp^{+i}(0,x^-,0)\,U_{0_\perp}[x^-,0]\,gF_\perp^{+i}(0,0,0)\,U_{0_\perp}[0,-\infty]
\right\}\right\rangle.
\nonumber\\
\label{eq:FF}
\end{eqnarray}
Note, however, that the cut-off cannot be set to infinity in lattice calculations, 
neither it is usually set to infinity in perturbative calculations \cite{Arnold:2008vd,CaronHuot:2008ni}. 

Finally, when doing perturbative calculations, it is useful to write the medium average as an exponential
\begin{equation}
\frac{1}{N_c}\left\langle\mathrm{Tr}\left\{[0,x_\perp]_-\, W^\dagger[0,x_\perp]\, [x_\perp,0]_+\, W[0,0] \right\}\right\rangle
= e^{C(x_\perp)L},
\qquad\text{for\;}L\rightarrow\infty\,.
\label{eq:defofcx}
\end{equation}
If one only considers the first order in the expansion of the
exponential, $e^{C(x_\perp)L} \approx 1 + C(x_\perp)L + \dots$, 
the Fourier transform of the quantity $C(x_\perp)$ is just the differential rate 
for elastic collisions of a quark with particles in the medium.
From \eqref{eq:qhat} and \eqref{eqn:complete3}, we thus have 
\begin{equation}
\hat{q}\approx\int_{k^2_\perp \leq q^{{\rm max}\,2}}\frac{d^2k_\perp}{(2\pi)^2}\,k_\perp^2\,C(k_\perp)\,,
\label{eqn:qhatC}
\end{equation}
where we have explicitly written the cut-off on the transverse momentum.
The left-hand side of eq.~\eqref{eq:defofcx} resembles very much 
a Wilson loop with a transverse extension $x_\perp$ stretching along the light-cone coordinate $x^-$, 
which, in the $x^+=0$ plane, is proportional to time, whereas $-C(x_\perp)$ resembles the corresponding static energy.
The analogy naturally leads to the use of lattice data from the static Wilson loop to determine 
$\hat{q}$. However, this requires some care, for the fields in the usual Wilson loop  
are time ordered \cite{Brown:1979ya}, while in the left-hand side of eq. \eqref{eq:defofcx} 
they are path ordered. We will discuss in the next section, in a special case, 
how this limitation may be circumvented and how we may use already 
existing lattice data to gain information on $\hat{q}$.

\section{\texorpdfstring{Application: contribution from the scale $g^2T$ to $\hat{q}$}{}}
\label{sec:ap}
As an example for an application of the gauge invariant formulation 
provided by eqs. \eqref{eq:defofcx} and \eqref{eqn:qhatC}, we consider 
the special case of a jet propagating in a weakly-coupled quark-gluon plasma at equilibrium. 
Because the plasma is weakly-coupled, it is characterized not only by the temperature,  
but also by a hierarchy of other energy scales: the Debye mass, $gT$, and 
the magnetic mass, $g^2T$. While one can compute perturbatively 
the contributions to $\hat{q}$ coming from the scales $T$ and $gT$,  
this is in general not possible for the scale $g^2T$, not even in the limit $g\to 0$ \cite{Gross:1980br}. 
Instead, a way to compute the contributions coming from the magnetic mass 
to the jet quenching parameter is using lattice gauge theories.
Following~\cite{Laine:2012he}, we will argue that the leading contribution coming from the magnetic mass 
is encoded in the static energy of an SU(3) Yang--Mills gauge theory in three dimensions,
and that it can be extracted from available lattice data. 

We proceed as follows. First, we introduce a cut-off $q^*$ to separate 
contributions coming from the momentum region $g^2T$ from contributions 
coming from higher-energy scales: $q^{\rm max} \gg gT \gg q^* \gg g^2T$.  
Then we observe that for momenta below the cut-off and up to corrections 
of relative order $g^2$, the ordering of the Wilson lines in the expression 
for $\hat{q}$ does not matter. The reason can be understood in the real-time 
formalism \cite{Landsman:1986uw,Thoma:2000dc} as due to the fact 
that, at momenta lower than $T$, the so-called symmetric propagator 
is the dominant contribution in all two gluon-field correlators, which is 
a consequence of the Bose-enhancement factor. 

At this point we make use of the analysis in \cite{CaronHuot:2008ni}. 
There, it was shown that correlators supported on space-like and light-like surfaces, 
like the left-hand side of eq. \eqref{eq:defofcx}, 
may be analytically continued from Minkowski to Euclidean space-time 
up to corrections of relative order $g^2$. After analytical continuation, 
we furthermore take advantage of the hierarchy of thermal scales by systematically 
integrating them out along the program first devised in \cite{Braaten:1994na}. 
Integrating out the temperature in the left-hand side of eq. \eqref{eq:defofcx} leads 
to a thermal field average in a three-dimensional EFT called electrostatic QCD (EQCD).
The only degrees of freedom of EQCD are the zero modes of the gluon fields, whereas 
all fermionic degrees of freedom and higher modes of the gluon fields have been 
integrated out. Integrating out the Debye mass leads to a thermal average in an 
EFT called magnetostatic QCD (MQCD); MQCD is a three-dimensional theory 
whose only degrees of freedom are the components $A^1$, $A^2$ and $A^3$ of the
gluonic field. In fact, at leading order in the coupling, MQCD is exactly 
an SU(3) Yang--Mills gauge theory in three Euclidean dimensions with coupling $g^2_{3D}=g^2T$.
It is precisely because the coupling provides the only dynamical scale of the theory  
that every Feynman diagram contributes to the same order, and, therefore, 
expectation values in MQCD that do not depend on some external larger scale 
cannot be evaluated in perturbation theory.

In MQCD, the left-hand side of \eqref{eq:defofcx} can be read by replacing 
the field $A^+$ with $A^3/\sqrt{2}$, since the field $A^0$ has been integrated out 
at the energy scale $g^2T$. Then, following~\cite{Laine:2012he}, 
we are in the position to relate the contribution to $C(x_\perp)$, 
coming from the magnetic mass, with the static energy in three-dimensional SU(3) gauge theory, $V(x_\perp)$:
\begin{equation}
C(x_\perp)\,\Big|_{g^2T} = -V(x_\perp)\,.
\label{eq:CV}
\end{equation}
The static energy $V(x_\perp)$ is a non-perturbative quantity at distances of order $1/(g^2T)$ 
that has been calculated on the lattice \cite{Luscher:2002qv}.  
Note that the identification \eqref{eq:CV} is possible only among gauge-invariant quantities, 
like $C(x_\perp)$ provided by (\ref{eq:defofcx}) is.  
In terms of $V(x_\perp)$, eq.~(\ref{eqn:qhatC}) can be rewritten as 
\begin{equation}
\hat{q}\,\Big|_{g^2T}=-(q^*)^2\int_0^\infty\,d\lambda\,\lambda^3\, 
J_0(\lambda)\int_\lambda^\infty\frac{dz}{z^3}\,V\left(\frac{z}{q^*}\right)\,,
\label{qhatV}
\end{equation}
where $J_0(\lambda)$ is the zeroth-order Bessel function. 
As suggested in \cite{Laine:2012he}, the lattice data for (the derivative of) $V(x_\perp)$ can be taken from \cite{Luscher:2002qv}
noticing that a constant shift in the potential would not contribute to \eqref{qhatV}.
The potential has a short-range tail that can be deduced from \cite{Pineda:2010mb};
a study hat accounts for some features of the short-range potential is given in~\cite{Laine:2012he}.
According to  \cite{Luscher:2002qv}, the long-range tail of the potential behaves like 
\begin{equation}
V(r)=\frac{1}{r_0}\left(a\,\frac{r}{r_0}-b\,\frac{r_0}{r} + \dots \right)\,,
\label{eq:vfit}
\end{equation}
for $r>r_0 \approx 2.2/g^2_{3D}$. The coefficient $a/r_0^2$ with $a \approx 1.5$ is the string tension and 
$b=\pi/24 \approx 0.13$ is the so-called L\"uscher term. By substituting \eqref{eq:vfit} into \eqref{qhatV}, we 
obtain for $\hat{q}$ the expression $\displaystyle a\,\frac{q^*}{r_0^2}+b\,\frac{(q^*)^3}{3} + \dots\,.$
This expression, if interpreted as an expansion in $q^*r_0$, is convergent for $q^*r_0 \siml 1$, reflecting 
the fact that it accounts for the long-range part of the potential only. 
A complete study, which is beyond the scope of the present work, would   
require the matching of the long-range part of the potential with a suitable short-range part.

An analysis along the above lines provides just the contribution from the region $k_\perp \sim g^2T$ that enters the
perturbative computation of $\hat{q}$ at NNLO, i.e. at order $g^6T^3$. 
In order to have a complete NNLO result, one would also need
the contributions from the regions $k_\perp \sim gT$ and $k_\perp \sim T$. 
What we would like to stress here, however, is that even in perturbation theory there are
contributions that have to be computed using lattice techniques. Such contributions require 
a fully gauge-invariant definition of $\hat{q}$, like the one derived in this work.
At present, $\hat{q}$ is known up to NLO in the $g$ expansion \cite{CaronHuot:2008ni} (see also \cite{D'Eramo:2011gs} for a 
leading-order analysis of eq. \eqref{eq:eramo} in different transverse momentum regimes), 
so non-perturbative physics will indeed be needed starting from the next order.

\section{Conclusions}
\label{sec:con}
We have derived a gauge invariant definition of the
jet quenching parameter $\hat{q}$ under the assumptions that the
medium is very large ($L\rightarrow\infty$) and that the jet energy, $Q$, 
is much larger than any other energy scale of the medium (e.g. $Q\gg T$). 
The existence of very different energy scales allows for the construction of an EFT, 
namely SCET supplemented with Glauber gluons, that describes the transverse momentum broadening 
of a highly energetic particle due to the medium. The effective theory is organized in a systematic expansion in $\lambda$, 
which is the small parameter associated with the ratio of the low energy scales of the medium and $Q$.
The specific power counting of the  SCET Lagrangian \eqref{eq:SCET} depends on the gauge:
some terms are enhanced in light-cone gauge with respect to a covariant gauge.
Hence, additional vertices need to be considered in the general gauge case with 
respect to the simpler covariant gauge case.
By a direct calculation of all diagrams containing these additional vertices 
at lowest order in $\lambda$, we have found that $\hat{q}$ is related to the
medium average of some previously known longitudinal Wilson lines as well as some transverse Wilson lines. 
The main result for the transverse momentum broadening probability is given in eq.~\eqref{eqn:complete}. 
The corresponding Wilson lines are shown in fig.~\ref{fig:p2}. 

With respect to other expressions that can be found in the literature, 
eq.~\eqref{eqn:complete}  appears to hold for all regularizations of the light-cone 
singularity in light-cone gauge. Moreover, it follows from an explicit 
resummation of Feynman diagrams and not from a heuristic extension of the 
covariant gauge result. The fields in the Wilson lines of \eqref{eqn:complete} are path ordered, 
which means that fields supported in one transverse plane are time ordered while
fields supported in the other transverse plane are anti-time ordered. 
This leads to some subtleties when explicitly proving the gauge invariance of the 
expression; a proof of gauge invariance can be found at the end of section \ref{sec:jball}.
Equivalent formulations of eq.~\eqref{eqn:complete} are given by eq.~\eqref{eqn:complete2} and 
eq.~\eqref{eqn:complete3}. The Wilson lines contributing to the latter are shown in  fig.~\ref{fig:p3}.

The fully gauge-invariant expression not only allows for computations in any gauge, 
but also opens the way for the use of lattice data in the evaluation of $\hat{q}$, 
as suggested in \cite{CaronHuot:2008ni,Majumder:2012sh,Laine:2012he}. 
A particularly suitable expression is provided by the transverse momentum broadening probability given in 
eq.~\eqref{eqn:complete3}, which translates into the expression of $\hat{q}$ given in 
eqs.~\eqref{eq:defofcx} and \eqref{eqn:qhatC}. As an example of using lattice data, we have discussed the contribution 
from the momentum region $k_\perp \sim g^2T$ to $\hat{q}$ in a weakly-coupled quark-gluon plasma.

The calculation of the jet broadening presented here includes the effect of Glauber and soft gluons. 
There are, however, other modes whose contributions may be relevant 
and that will need to be considered before comparing with data. 
The addition of these new modes, in particular collinear modes \cite{D'Eramo:2011zz}, 
may be systematically accounted for in the SCET framework.
Also the calculation of $\hat{q}$ in a weakly-coupled quark-gluon plasma may be improved 
by including the remaining (perturbatively calculable) NNLO terms~\cite{Benzke2012}.

\paragraph{Acknowledgments}

We thank Sean Fleming, Ahmad Idilbi, Abhijit Majumder, Grigory Ovanesyan and Antonio Pineda for useful discussions.
We are grateful to Mikko Laine for discussions and comments on section 4. 
The support by the Excellence Cluster ,,Origin and Structure of the Universe'' is gratefully acknowledged. 
This research is supported by the DFG grants BR 4058/2-1 and BR 4058/1-1.

\appendix

\section{\texorpdfstring{Computation of $G^\pm_n(q)$}{Computation of G+-n(q)}}
\label{sec:ape-}
In this appendix, we derive eqs. (\ref{eq:res-}) and \eqref{eq:res1-}. 
The proof of eqs. (\ref{eq:res+}) and (\ref{eq:res1+}) is analogous and, for this reason, 
will not be detailed here.

We will proceed as follows. First, we prove  eqs. \eqref{eq:res-} and \eqref{eq:res1-} 
for the cases $n=1$ and $n=2$. After that, we will show that if the relations
(\ref{eq:res-}) and (\ref{eq:res1-}) are fulfilled for $n-2$ and $n-1$, then they are fulfilled for
$n$, which proves eqs. (\ref{eq:res-}) and \eqref{eq:res1-} by induction.

\subsection{\texorpdfstring{Computation of $G^-_1(q)$}{}}
The computation of $G^-_1(q)$ is straightforward; by just writing down the Feynman rule we obtain
\begin{equation}
G_1^-(q)=-\frac{iq_\perp\cdot gA_\perp^{\rm sin}(q-q_0)}{2Q}\,\sh{n}\,.
\end{equation}
Expressing the gluon field in position space and using eq. \eqref{eqn:puregauge}, we can then write
\begin{equation}
G_1^-(q)=-\frac{1}{2Q}\int\,dy^+\,dy^-\theta(-y^-)\,e^{i(q^--Q)y^++iq^+y^-}q_\perp^2\;g\phi^-(y^+,q_\perp)\sh{n}\,.
\label{appg1}
\end{equation}
Comparing with eq. (\ref{eq:res-}), eq. \eqref{appg1} implies
\begin{equation}
f_1(y^+,q_\perp)=-\frac{q_\perp^2}{2Q}\;g\phi^-(y^+,q_\perp)\,,
\end{equation}
which agrees with \eqref{eq:res1-} for the case $n=1$.

\subsection{\texorpdfstring{Computation of $G_2^-(q)$}{}}
From eq. \eqref{eq:rel-}, it follows that $G_2^-(q)$ satisfies
\begin{equation}
G_2^-(q)=\raisebox{-28pt}{\includegraphics[width=32mm]{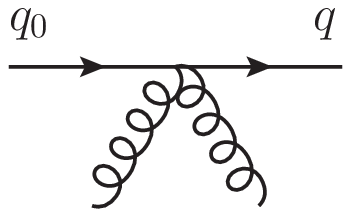}}
+\int\frac{d^4q_1}{(2\pi)^4}\,G^-_1(q_1)\times \raisebox{-30pt}{\includegraphics[width=45mm]{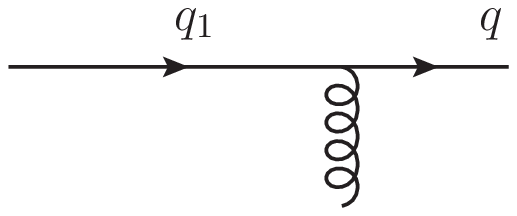}}\,.
\end{equation}
Hence, $G_2^-(q)$ gets a 2-gluon vertex contribution, namely
\begin{equation}
G^-_{2a}(q)=\raisebox{-28pt}{\includegraphics[width=32mm]{./G2m2g.eps}}\,,
\end{equation}
and a 1-gluon vertex contribution, 
\begin{equation}
G^-_{2b}(q)=\int\frac{d^4q_1}{(2\pi)^4}\,G^-_1(q_1)\times\raisebox{-30pt}{\includegraphics[width=45mm]{./G1m1g.eps}}\,.
\end{equation}
Computing the two, we obtain 
\begin{align}
G^-_{2a}(q)=\frac{i}{2Q}\int\frac{d^2{q_{1\,\perp}}}{(2\pi)^2}\,dy^+\,dy^-\theta(-y^-)&\,e^{i(q^--Q)y^++iq^+y^-}
\nonumber\\
&\hspace{-10mm}
\times \left({q}_\perp-{q_{1\,\perp}}\right)\cdot{q_{1\,\perp}}\;g\phi^-(y^+,q_\perp-{q_{1\,\perp}})\; g\phi^-(y^+,{q_{1\,\perp}})\sh{n}\,,
\\
G^-_{2b}(q)=-\frac{i}{2Q}\int\frac{d^2{q_{1\,\perp}}}{(2\pi)^2}\,dy^+\,dy^-\theta(-y^-)&\,e^{i(q^--Q)y^++iq^+y^-}
\nonumber\\
&\hspace{-10mm}
\times\left(q^2_\perp-{q^2_{1\,\perp}}\right)\;g\phi^-(y^+,q_\perp-{q_{1\,\perp}})\; g\phi^-(y^+,{q_{1\,\perp}})\sh{n}\,,
\end{align}
and finally $G^-_2(q)$, which is of the form (\ref{eq:res-}), but with $f_2$ given by
\begin{eqnarray}
f_2(y^+,q_\perp) &=&\frac{i}{2Q}\int\frac{d^2{q_{1\,\perp}}}{(2\pi)^2}\,
\left[(q_\perp-{q_{1\,\perp}})\cdot{q_{1\,\perp}}-\left(q^2_\perp-{q^2_{1\,\perp}}\right)\right]
\nonumber\\
&& \hspace{40mm}
\times \;g\phi^-(y^+,q_\perp-{q_{1\,\perp}})\;g\phi^-(y^+,{q_{1\,\perp}})\,.
\label{eq:f2}
\end{eqnarray}
In order to show that eq. \eqref{eq:f2} is equivalent to \eqref{eq:res1-}, we 
express $\phi^-$ in position space and use eq. \eqref{derPphi}:
\begin{eqnarray}
f_2(y^+,q_\perp) &=& \frac{i}{2Q}\int\frac{d^2{q_{1\,\perp}}}{(2\pi)^2}\,q_\perp\cdot\left( {q_{1\,\perp}} - {q}_\perp \right)
\;g\phi^-(y^+,q_\perp-{q_{1\,\perp}}) \; g\phi^-(y^+,{q_{1\,\perp}})
\nonumber\\
&=& \frac{i}{2Q}\int d^2y_\perp\, d^2x_\perp\,\frac{d^2{q_{1\,\perp}}}{(2\pi)^2}\,
q_\perp\cdot\left( -i\nabla_\perp e^{-i(q_\perp-{q_{1\,\perp}})\cdot y_\perp} \right) e^{-i{q_{1\,\perp}}\cdot x_\perp}
\nonumber\\
&& \hspace{50mm}
\times\;g\phi^-(y^+,y_\perp) \;g\phi^-(y^+,x_\perp)
\nonumber\\
&=& -\frac{i}{2Q}\int d^2y_\perp\, e^{-iq_\perp\cdot y_\perp} q_\perp\cdot \left( -i\nabla_\perp \, g\phi^-(y^+,y_\perp) \right) 
\;g\phi^-(y^+,y_\perp)
\nonumber\\
&=&\frac{i}{4Q}\, q_\perp^2\int d^2y_\perp\,e^{-iq_\perp\cdot y_\perp}\,
{\rm P}\left(\left[ig\phi^-(y^+,y_\perp)\right]^2\right)\,,
\label{eqn:f2}
\end{eqnarray}
which agrees with \eqref{eq:res1-} for the case $n=2$.

\subsection{Proof by induction}
In order to complete the proof by induction, we need to show that if $G^-_{n-1}$ and $G^-_{n-2}$ fulfill 
eqs. (\ref{eq:res-}) and \eqref{eq:res1-}, then $G^-_n$, defined through eq. (\ref{eq:rel-}),   
also fulfills them. Specifically, from eq. (\ref{eq:rel-}) it follows that $G^-_n$ is of the form 
(\ref{eq:res-}) with $f_n$ given by 
\begin{align}
f_n(y^+,q_\perp)=&
i\int\frac{d^2{q_{n-1\,\perp}}}{(2\pi)^2}\,\frac{q^2_\perp-{q^2_{{n-1}\,\perp}}}{q^2_{{n-1}\,\perp}}
\;g\phi^-(y^+,q_\perp-{q_{n-1\,\perp}}) \;f_{n-1}(y^+,{q_{n-1\,\perp}})
\nonumber\\ 
&+\int\frac{d^2{q_{{n-1}\,\perp}}}{(2\pi)^2}\,\frac{d^2{q_{n-2\,\perp}}}{(2\pi)^2}
\,\frac{\left(q_\perp-{q_{n-1\,\perp}}\right)\cdot\left({q_{n-1\,\perp}}-{q_{n-2\,\perp}}\right)}{{q^2_{n-2\,\perp}}}\,
\nonumber\\
&\times\; g\phi^-(y^+,q_\perp-{q_{n-1\,\perp}}) \;g\phi^-(y^+,{q_{n-1\,\perp}}-{q_{n-2\,\perp}})\;f_{n-2}(y^+,{q_{n-2\,\perp}})\,.
\end{align}
Using the expressions of $f_{n-1}$ and $f_{n-2}$ given in eq. (\ref{eq:res1-}), we get
\begin{align}
f_n(y^+,q_\perp)=& - \frac{1}{2Qn!}
\int\frac{d^2{q_{n-1\,\perp}}}{(2\pi)^2}\,d^2y_\perp\, 
n\,\left(q^2_\perp-{q^2_{n-1\,\perp}}\right)
\;g\phi^-(y^+,q_\perp-{q_{n-1\,\perp}})\; 
\nonumber\\
&\hspace{30mm}
\times 
e^{-i{q_{n-1\,\perp}}\cdot y_\perp} 
{\rm P}\left(\left[ig\phi^-(y^+,y_\perp)\right]^{n-1}\right)
\nonumber\\
& + \frac{i}{2Qn!}
\int\frac{d^2{q_{n-1\,\perp}}}{(2\pi)^2}\,\frac{d^2{q_{n-2\,\perp}}}{(2\pi)^2}\,d^2y_\perp\, 
n(n-1)\,\left(q_\perp-{q_{n-1\,\perp}}\right)\left({q_{n-1\,\perp}}-{q_{n-2\,\perp}}\right)
\nonumber\\
&\hspace{30mm}
\times\;g\phi^-(y^+,q_\perp-{q_{n-1\,\perp}})\;g\phi^-(y^+,{q_{n-1\,\perp}}-{q_{n-2\,\perp}}) 
\nonumber\\
&\hspace{30mm}
\times e^{-i{q_{n-2\,\perp}}\cdot y_\perp} 
{\rm P}\left(\left[ig\phi^-(y^+,y_\perp)\right]^{n-2}\right)\,.
\end{align}
Finally, by using similar manipulations as in (\ref{eqn:f2}), this can be brought into the form of eq.~\eqref{eq:res1-}.

\end{document}